\documentclass{aastex631}

\newcommand{\DM}{{\rm DM}}

\usepackage{CJK}
\usepackage{physics}
\usepackage{array}
\usepackage{hyperref}

\newcommand{\fdiff}{0.84}

\newcommand{\SMCMC}{0.133_{-0.045}^{+0.034}}
\newcommand{\HOf}{2.813_{-0.258}^{+0.250}}
\newcommand{\Sigmahost}{0.605_{-0.134}^{+0.173}}
\newcommand{\Expmu}{182.937_{-42.032}^{+43.943}}

\newcommand{\KsigmaS}{0.979}
\newcommand{\LogKsigmaS}{-0.00928}

\newcommand{\F}{0.357^{+0.043}_{-0.067}}
\newcommand{\Hubble}{66.889_{-5.459}^{+6.754}}

\newcommand{\MacS}{0.441^{+0.055}_{-0.096}}
\newcommand{\MacScut}{0.409^{+0.579}_{-0.270}}

\begin{document}
\begin{CJK*}{UTF8}{gbsn}
\title{
Hubble constant constraint using 117 FRBs with a more accurate probability density function for $\DM_{\rm diff}$
}

\author[0000-0001-8114-3094]{Jiaming Zhuge(诸葛佳明)}\thanks{jiaming.zhuge@unlv.edu}
\affiliation{Nevada Center for Astrophysics, University of Nevada, Las Vegas, NV 89154, USA}
\affiliation{Department of Physics and Astronomy, University of Nevada, Las Vegas, NV 89154, USA}

\author[0000-0001-6677-949X]{Marios Kalomenopoulos} \thanks{marios.kalomenopoulos@gmail.com}
\affiliation{Nevada Center for Astrophysics, University of Nevada, Las Vegas, NV 89154, USA}
\affiliation{Department of Physics and Astronomy, University of Nevada, Las Vegas, NV 89154, USA}

\author[0000-0002-9725-2524]{Bing Zhang(张冰)}\thanks{bing.zhang@unlv.edu}
\affiliation{Nevada Center for Astrophysics, University of Nevada, Las Vegas, NV 89154, USA}
\affiliation{Department of Physics and Astronomy, University of Nevada, Las Vegas, NV 89154, USA}

\begin{abstract}

Fast radio bursts (FRBs) are among the most mysterious astronomical transients. Due to their short durations and cosmological distances, their dispersion measure (DM) - redshift ($z$) relation is useful for constraining cosmological parameters and detecting the baryons in the Universe. The increasing number of localized FRBs in recent years has provided more precise constraints on these parameters. However, the larger dataset reveals limitations in the widely used probability density function ($p_{\rm diff}$) for $\DM_{\rm diff}$, which refers to the diffuse electron term of FRB DM. 
In this project, we collect 117 of the latest, localized FRBs, discuss the effect of a more accurate $\sigma_{\rm diff}$, which is a parameter in $p_{\rm diff}$ and once thoughts as ``effective standard deviation'', and more clearly rewrite their likelihood to better constrain the parameters above. We find that the widely used approximation $\sigma_{\rm diff} \sim F/\sqrt{z}$ only works under contrived assumptions and shows the greatest deviation from the true standard deviation in low redshift. In general, one should use an accurate method to derive this parameter from $p_{\rm diff}$. Our method yields better constraints on $H_0\Omega_b f_{\rm diff} = \HOf\;{\rm km/s/Mpc}$ or $H_0 = \Hubble \;{\rm km/s/Mpc}$ when combining the FRB data with CMB measurements and taking $f_{\rm diff} = 0.84$. This fully analytical correction helps us better constrain cosmological parameters with the increasing number of localized FRBs available today.

\end{abstract}


%%%%%%%%%%%%%%%%%%%%%%%%%%%%%%%%%%Example%%%%%%%%%%%%%%%%%%%%%%%%%%%%%%%%%%%%%%

% \footnote{\url{http://www.latex-project.org/}}
% \citep{lamport94}
% \footnote{see Section \ref{sec:pubcharge}
% \url{http://journals.aas.org/authors/aastex.html}

\section{Introduction} \label{sec:intro}

Fast radio bursts (FRBs) are millionsecond duration bursts first discovered by \citet{2007Lorimer}. Predominantly located at cosmological distances, these radio bursts have very high brightness temperatures and luminosities \citep{2013Thornton_FRB_cos}. Their dispersion measure (DM)-redshift dependence (Section \ref{sec:DM}) provides an independent way to study cosmological parameters, as well as the cosmic baryon distribution. \citep[for reviews on FRBs, see e.g.][]{2018Katz_FRB, Popov_2018, 2019Cordes_FRB, Petroff2019, PLATTS2019, 2020Nature_Zhang, Xiao2021, 2022Xiao, 2022Petroff_FRB, 2023Zhang_phy_of_FRB, 2024Zhang_FRB_counterparts} 

The early idea of using FRB DM to constrain cosmological parameters can be traced back to \citet{2014ApJ_Deng, Gao_2014, 2014_BeiZhou} (see also earlier discussion of DM-$z$ relation of cosmological sources in general \citep{Ioka_2003,inoue_2004}), who only considered possible association events (for example, FRBs possibly associated with GRB events) or mock events due to the lack of localized FRBs. The breakthrough happened in 2020, when \citet{2020Nature_Macquart} used 5 localized FRBs, their ``golden sample'', together with physically motivated $\DM$ PDFs, to constrain cosmological parameters and astrophysical properties of the FRB sources. 

With the increasing number of localized FRBs, FRB cosmology has become a popular topic in transient astronomy \citep[e.g.][]{2024ApJ_Jay_F_H0, 2024_Surajit, Connor_Sharma, Sharma_2025_ApJ, 2025ApJ_Yiying_Wang_DESI, 2025ApJ_117, FRB_2025_Acharya}. However, cosmological constraints from the new, localized FRBs always show much larger values of the parameter $F$. This parameter was once thought to quantify the effective standard deviation through the relation $\sigma_{\rm diff}\sim F/\sqrt{z}$, where $\sigma_{\rm diff}$ is a parameter in the probability density function ($p_{\rm diff}$) of the diffuse electron dispersion measure ($\DM_{\rm diff}$)(see the detailed discussion in Section \ref{subsec:diff_PDF}). This problem is urgent and has attracted attention in the community. Results from \citet{2021_Zijian_Zhang} implied that the power of $z$ in $\sigma_{\rm diff}\sim F/\sqrt{z}$ may be smaller according to the TNG300 dataset, simulated from IllustrisTNG \citep{2019_illustrisTNG}. \citet{2024ApJ_Jay_F_H0} found that the IllustrisTNG simulations tend to underpredict $F$ for small redshifts ($z<0.4$), and importantly showed that the parameter is degenerate with $H_0$. \citet{Connor_Sharma} argued that the $F$ parameterization fails to adequately consider the contribution of IGM structure, such as the intersection of filaments and voids, while \citet{Sharma_2025_ApJ} quantified the inaccuracies of the Macquart PDF by analyzing hydrodynamical simulations with different feedback strengths. They found that the $F$ parameter misestimates the variance of the PDF in the redshift range $0<z<2$, and eventually they proposed a new log-normal PDF for modeling the $\DM_{\rm diff}$.

Cosmological simulations also call for modifications in the PDF, but the results are under debate. \citet{2021MNRAS_TR_cosmology_simulation_TNG} first compared different mass resolution runs of the TNG300 dataset, comparing simulated DM statistics with analytical ones. They found that the standard deviation of the $\DM$ calculated from the simulations $\sigma_{\DM}$ over the square-root of redshift remains constant with time, i.e., $\sigma_{\DM}/\sqrt{z} \sim 230 {\rm pc/cm^3}$. \citet{2025ApJ_MI_cosmology_simulation_all_compare} studied, among other things, the evolution of the $F$ parameter with time across different simulations—IllustrisTNG \citep{2019_illustrisTNG}, SIMBA \citep{SIMBA}, and Astrid \citep{Astrid} from the CAMELS \citep{CAMELS} suite—observing it remains largely constant across the redshift range $z = 0-2$. This is in contrast with \citet{2021_Zijian_Zhang}, who found that $F$ is diverging for low-z in the IllustrisTNG simulation. \citet{2024arXiv_ZJ_cosmology_simulation_CoDa} provided a new fitting result from CoDa II \citep{CoDa2}, where they investigated the evolution of $\DM$ and its standard deviation $\sigma_{\DM}$ to higher redshifts ($z \leq 12$). 

In this paper, we derive the parameter $\sigma_{\rm diff}$ in $p_{\rm diff}$ from the true standard deviation to get more precise result in Section \ref{subsec:diff_PDF}, based only on more accurate equations analytically. We also show the FRB $\DM$ likelihood in Section \ref{subsec:likelihood}. These two consideration help us to get better constraints. We first review the idea of FRB cosmology in Section \ref{sec:DM}. In Section \ref{sec:data}, we introduce our 117 latest localized FRBs' dataset and how we preprocess them. The key points of this paper are presented in Section \ref{sec:met}, where we show the PDFs used in this work and how we consider the critical parameters $\sigma_{\rm diff}$. We also derive a clear form of the modified convolution to calculate the likelihood. In Section \ref{sec:results}, we show all of our novel constraints, and the conclusions are summarized in Section \ref{sec:conclusion}.

\section{From DM to cosmology}\label{sec:DM}

The DM is connected to the difference of the arrival times of the electromagnetic waves in different frequencies when the waves propagate through an ionized gas. For two photons traveling from a cosmological distance of redshift $z$, with frequencies $\nu_1$ and $\nu_2$ much greater than the plasma frequency $\omega_p$, their arrival time difference in the observer frame is \citep{1979Radiative_processes_in_astrophysics, 2014ApJ_Deng}:
\begin{equation}
    \Delta t=\frac{e^2}{2\pi m c}\left(\frac{1}{\nu_1^2}-\frac{1}{\nu_2^2}\right)\int_0^L \frac{n_e(z)}{1+z}\dd{l}.
\end{equation}
Here, $n_e$ is the number density of electrons, $L$ is the proper distance from the FRB source to the observer. The dispersion measure is defined as the final integration term:
\begin{equation}
    \DM=\int_0^L \frac{n_e(z)}{1+z}\dd{l}.
\end{equation}

The DM of an FRB is composed of the following three components ordered from local to distant \citep[e.g.][]{2013Thornton_FRB_cos, 2014ApJ_Deng, Gao_2014, 2019_Prochaska_Zheng, James2021}:
\begin{equation}
    {\DM_{\rm FRB}}={\DM_{\rm MW}}+{\DM_{\rm diff}}+\frac{{\DM_{\rm host}}}{1+z}.
    \label{eq:FRB_DM}
\end{equation}
Here, $\DM_{\rm MW}$ is the Milky Way component, which consists of ${\DM_{\rm MW, ISM}}$ and ${\DM_{\rm MW, halo}}$, which are the interstellar medium and the Milky Way halo terms, respectively. The term ${\DM_{\rm MW, ISM}}$ can be calculated with the Galactic electron models such as NE2001 \citep{2002_NE2001I, 2003_NE2001II} and YMW16 \citep{2017_YMW16}, which are derived from pulsar data. ${\DM_{\rm diff}}$ is the contribution of diffuse electrons, which is the sum of $\DM$ in the intergalactic medium (IGM) (${\DM_{\rm IGM}}$) and $\DM$ in cosmic halos (${\DM_{\rm halo}}$). ${\DM_{\rm host}}$ is the host environment term, which includes the contribution from the FRB's host galaxy and the immediate source environment (${\DM_{\rm src}}$). Note that ${\DM_{\rm host}}$ is corrected by a factor of $(1+z)$ due to its cosmological distance. 

${\DM_{\rm diff}}$ is the core component when inferring cosmology and the baryon distribution in the Universe. In $\Lambda {\rm CDM}$ cosmology, the mean value of the dispersion measure of diffuse matter can be written as \citep{2014ApJ_Deng, Gao_2014, 2014_BeiZhou, 2020Nature_Macquart}:
\begin{equation}
    \langle{\DM_{\rm diff}}\rangle 
    (z) = \frac{3cH_0\Omega_b }{8\pi G m_p}\int_0^z \frac{ f_{\rm diff}(z')\chi(z')(1+z')}{\sqrt{\Omega_m(1+z')^3+\Omega_{\Lambda}}}\dd{z'},
    \label{eq:DM_diff}
\end{equation}
with
\begin{equation}
    \chi(z)=Y_H X_{e,H}(z)+\frac{1}{2}Y_p X_{e,He}(z)
    \label{eq:chi}
\end{equation} 
We take $\langle{\DM_{\rm diff}}\rangle$ to identify the theoretical $\DM_{\rm diff}$. Here, $\Omega_b$ is the dimensionless baryon mass fraction of the Universe, $f_{\rm diff}$ is the fraction of the diffuse baryons in both IGM and halos, $\chi(z)$ is the fraction of ionized electrons to baryons and can be expanded as in Eq. \ref{eq:chi}, $Y$ is the mass fraction of hydrogen or helium, and $X$ is the ionization fraction for each element. In the low-redshift universe ($z<3$), both H and He are fully ionized, i.e. $\chi(z)\sim 7/8$ \citep{2014ApJ_Deng}.  

$\DM_{\rm diff}$ is a function of redshift. It is related to the baryon distribution $f_{\rm diff}$, the ionized fraction $\chi$ and the cosmological parameters $H_0, \Omega_b$. This means that additionally to the comic microwave background or type-Ia supernovae, the $\DM_{\rm diff}$ is another cosmic probe to study the different fractions of the components in the universe, as well as its reionization history. 

\section{Data} \label{sec:data}

In this project we collect the 117 latest localized FRBs with their redshifts and DM (Table \ref{tab:data}) \citep[e.g.][]{Blinkverse, Connor_Sharma, 2025ASKAP, highres_ASKAP_arxiv2025}. We deduct the term $\DM_{\rm MW, ISM}$ using the NE2001 model \citep{2002_NE2001I, 2003_NE2001II}, each calculated individually for each FRB based on its sky localization. ${\DM_{\rm MW, halo}}$, according to previous studies, ranges from $30 \, {\rm pc/cm^3}$ to $80 \, {\rm pc/cm^3}$ \citep[][]{DM_Xu_2015, DM_Dolag, 2019_Prochaska_Zheng, DM_Yamasaki_2020, DM_Arcus, Hashimoto_2020}. In this project we take ${\DM_{\rm MW, halo}}\sim 30 \, {\rm pc/cm^3}$ for two reasons. 
First, a small ${\DM_{\rm MW, halo}}$ will result in more positive $\DM_{\rm ext}$ and retain more data with physical $\DM$ values to constrain. Second, the add-on value of ${\DM_{\rm MW, halo}}$ can be absorbed in the other DM components, if our assumed value proves too small.
The standard deviations in ${\DM_{\rm MW, halo}}$ and ${\DM_{\rm MW, ISM}}$ will be absorbed in the standard deviation of ${\DM_{\rm host}}$. Thus, we can define the extragalactic $\DM_{\rm ext}$ as the observed DM ($\DM_{\rm obs}$) subtracting the MW term, such that $\DM_{\rm ext}$ includes the diffuse electron term and host environment term:
\begin{equation}
    \DM_{\rm ext}=\DM_{\rm obs}-\DM_{\rm MW, halo}-\DM_{\rm MW, ISM}=\DM_{\rm diff}+\frac{\DM_{\rm host}}{1+z}
\end{equation}

\begin{figure}
\centering
\includegraphics[width=0.7\textwidth]{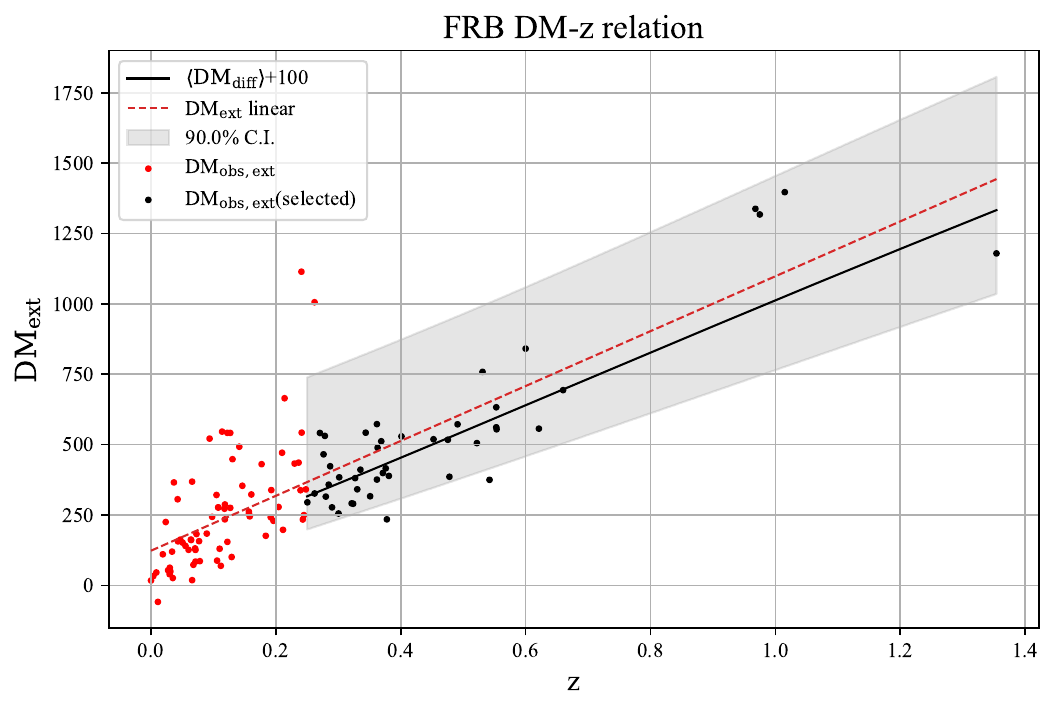}
\caption{$\DM_{\rm ext}-z$ relation for 117 localized FRBs, where $\DM_{\rm ext}=\DM-\DM_{\rm MW}$ is the extragalactic DM. The dashed red line is the best fitting linear function for all data, while the back line shows the theoretical line from Eq. \ref{eq:DM_diff} adding $100\,\rm pc\,cm^{-3}$ for $\DM_{\rm host}$. We also show 90$\%$ confidence interval in shaded region calculated from our parameters searching results. The black data points are those FRBs $z\ge 0.25$ that $p_{\rm diff}(\Delta)$ worked within this range.}
\label{fig:DM_z}
\end{figure}

For all FRBs, a linear least squares fit yields:
\begin{equation}
    {\DM_{\rm ext}}={\DM-\DM_{\rm MW}}=(975.27z+ 123.06 )\ {\rm pc/cm^3}.
\end{equation}
The slope gives us a rough estimate for the parameters in $\langle{\rm DM_{\rm diff}}\rangle (z)$, and the intercept reflects the value of ${\DM_{\rm host}}$ \citep[e.g.][]{2023Zhang_phy_of_FRB}. Our FRBs data sample and the linear fit function are shown in Fig. \ref{fig:DM_z}. 

To prevent ${\DM_{\rm src}}$ from being the dominant term, we drop two special FRBs, FRB 20190520B and FRB 20220831A, which have peculiarly high ${\DM_{\rm ext}}$ but relatively small redshift. We also apply the redshift constraint $z\geq0.25$, retaining 44 FRBs. This redshift cut is determined from our analysis of the PDF $p_{\rm diff}$ limits (see details in Section \ref{subsubsection:sigma_diff}).

\section{Methods} \label{sec:met}

After correcting for MW effects, the only two terms left are $\DM_{\rm diff}$ and $\DM_{\rm host}$. Here, we use the maximum likelihood method and Markov Chain Monte Carlo (MCMC) to study the parameters in the PDFs described below and constrain cosmology. 

% \subsection{Probability density function}

\subsection{PDF for Host galaxy term}

For ${\DM_{\rm host}}$, we take a log-normal distribution:
\begin{equation}
    p_{\rm host}({\DM_{\rm host}|\mu, \sigma_{\rm host}})=\frac{1}{\sqrt{2\pi}\sigma_{\rm host}{\DM_{\rm host}}}\exp[-\frac{(\ln{\DM_{\rm host}-\mu})^2}{2\sigma_{\rm host}^2}],
    \label{eq:p_host}
\end{equation}
which has an asymmetric tail in the large value side. The asymmetric tail allows for the existence of a large value ${\DM_{\rm host}}$, which may come from a dense local environment. The median value for this function is $\exp(\mu)$. The variance of this function is $[\exp(\sigma_{\rm host}^2)-1]\exp(2\mu+\sigma_{\rm host}^2)$. We take $\exp(\mu)$ and $\sigma_{\rm host}$ as two unknown parameters to search for. Note that the variable $\DM_{\rm host}$ here should be the value in the comoving frame because in Eq. \ref{eq:FRB_DM} we divide $\DM_{\rm host}$ by $(1+z)$.

\subsection{PDF for Diffuse electron term}\label{subsec:diff_PDF}

For ${\DM_{\rm diff}}$, we take the distribution:
\begin{equation}
    p_\Delta(\Delta)={\cal A} \Delta^{-\beta}\exp\left[-\frac{(\Delta^{-\alpha}-C_0)^2}{2\alpha^2\sigma_{\rm diff}^2}\right],
    \label{eq:p_diff}
\end{equation}
which was first used in \citep{McQuinn_2014, 2019_Prochaska_Zheng, 2020Nature_Macquart} and was derived from modeling both observational and simulation data. Here, $\Delta$ refers to the dimensionless fraction between the measured and theoretical diffuse DM terms, which is $\Delta={\DM_{\rm diff}}/\langle{\DM_{\rm diff}}\rangle$; ${\cal A}$ is the normalization factor, calculated imposing $\int p_\Delta d\Delta =1$; $C_0$ is a factor that affects the mean value (Since this PDF is highly right-skewed, to find $C_0$ we impose the condition that the mean value of $\Delta$ equals to $1$, which holds by definition $\langle{\Delta}\rangle = \langle{\DM_{\rm diff}}\rangle/\langle{\DM_{\rm diff}}\rangle=1$); $\alpha$ and $\beta$ are the parameters that describe the gas profile in cosmic halos and we adopt $\alpha=\beta=3$ following \citet{2020Nature_Macquart}; Finally $\sigma_{\rm diff}$ is related to the scatter of $\Delta$ and is redshift-dependent. 

\subsubsection{A more accurate $\sigma_{\rm diff}$ and the limitation of $F$}\label{subsubsection:sigma_diff}

$\sigma_{\rm diff}\sim F/\sqrt{z}$ was once thought to be the ``effective standard deviation''\citep{2020Nature_Macquart}. However, in this PDF form, $\sigma_{\rm diff}$ may deviate greatly from the true standard deviation of $\Delta$. To prevent any confusion, we define a new variable $\sigma_\Delta$ as the true standard deviation to differentiate it from $\sigma_{\rm diff}$. An easy way to estimate $\sigma_\Delta$ is to consider its contributions from the intergalactic medium and halos, since $\DM_{\rm diff}=\DM_{\rm IGM}+\DM_{\rm halos}$. Because $\Delta$ is $\DM_{\rm diff}$ normalized by its theoretical result $\langle{\DM_{\rm diff}}\rangle$, the standard deviation of $\Delta$ can be written as ${\sigma_\Delta}\sim\sqrt{(\sigma^2_{\rm IGM}+\sum_i \sigma^2_{\rm halo,i})/{\langle{\DM_{\rm diff}}\rangle^2}}$. Here, $\sigma_{\rm halos, i}$ stands for the standard deviation contribution of each halo along the line of sight from the FRB source to the Milky Way. For sufficiently large distances, the sum of $\sigma_{\rm halos, i}$ will dominate the standard deviation (the IGM-dominated case is shown in Eq. \ref{eq:var_delta_IGM}). To simplify the quantity, we define their mean square as
\begin{equation}
    \overline{\sigma^2}_{\rm halo} \sim 1/N \cdot \sum_i \sigma^2_{\rm halo,i},
\end{equation}
where $N$ is the total number of halos along the line of sight. At a substantial distance, $\DM_{\rm halos}$ is approximately equal to the product of the number density of halos $n_{\rm halo}$ (in the comoving volume), the comoving distance to the source of the FRB $D_c$, and the typical cross section of the halo $A_{\rm halo}$. One can define a ``mean observed halo path'' as $l\sim1/(n_{\rm halo}A_{\rm halo})$. Thus the number of halos in the line of sight, $N\sim n_{\rm halo} A_{\rm halo} D_c\sim D_c/l$,  is approximately the distance to the FRB source over the typical distance for one halo, and has a substantial $\DM$ standard deviation contribution with
\begin{equation}
    {\sigma_\Delta}\sim
    \sqrt{\frac{\sigma^2_{\rm IGM}+\sum_i \sigma^2_{\rm halo,i}}
    {\langle{\DM_{\rm diff}}\rangle^2}}\sim
    \sqrt{\frac{N\overline{\sigma^2}_{\rm halo}}
    {\langle{\DM_{\rm diff}}\rangle^2}}\sim
    \sqrt{\frac{\overline{\sigma^2}_{\rm halo}n_{\rm halo} A_{\rm halo} D_c}
    {\langle{\DM_{\rm diff}}\rangle^2}}\sim
    \sqrt{\frac{\overline{\sigma^2}_{\rm halo}D_c}{ 
    {\langle{\DM_{\rm diff}}\rangle^2 l}}}.
    \label{eq:var_delta}
\end{equation}
We can further expand this result by recalling Eq. \ref{eq:DM_diff} and the definition of comoving distance $D_c=c/H_0\int_0^z \dd{z'}/E(z')$ \citep{cos_dis}. For convenience, we define $E(z)=\sqrt{\Omega_m(1+z)^3+\Omega_{\Lambda}}$ (in $\Lambda$CDM cosmology). As for $l$, we can take $n_{\rm halo} \sim 10^{-3}\, {\rm Mpc}^{-3}$ and $A_{\rm halo}\sim 1 \,{\rm Mpc}^2$ to get an order of magnitude estimation $l\sim1/(n_{\rm halo}A_{\rm halo})\sim 1{\rm Gpc}$ \citep{1974ApJ_halo, Longair_galform_2023, 2025_ZhaoZhang_CROCODILE}. If we write $\sigma_\Delta$ as a function of the redshift $z$, we can absorb all constants into a parameter $S$ to get 
\begin{align}
    {\sigma_\Delta}
    &\sim\frac{64\pi G m_p}{21 \sqrt{c}}\sqrt{\frac{\overline{\sigma^2}_{\rm halo}}{H^{3}_0\Omega^2_b f^2_{\rm diff}l}}\frac{\sqrt{\int_0^z \dd{z'}/E(z')}}{\int_0^z (1+z')\dd{z'}/E(z')}\\
    \label{eq:var_delta_value}
    &\sim \sqrt{0.115}
    \left(\frac{H_0}{67\rm \,km\,s^{-1}\,Mpc^{-1}}\right)^{-\frac{3}{2}}
    \left(\frac{\Omega_b}{0.04897}\right)^{-1}
    \left(\frac{f_{\rm diff}}{0.84}\right)^{-1}
    \left(\frac{\overline{\sigma^2}_{\rm halo}}{(130\ \rm pc\,cm^{-3})^2}\right)^{\frac{1}{2}}
    \left(\frac{l}{1 \rm Gpc}\right)^{-\frac{1}{2}}
    \frac{\sqrt{\int_0^z \dd{z'}/E(z')}}{\int_0^z (1+z')\dd{z}/E(z')}\\
    \label{eq:var_delta_S}
    &\sim \frac{\sqrt{S\int_0^z \dd{z'}/E(z')}}{\int_0^z (1+z')\dd{z}/E(z')}.
\end{align}
Note that the values $n_{\rm halo}$ and $A_{\rm halo}$ used here are only rough estimates, as they are actually functions of mass and redshift. The realistic values for $l$ should be obtained by integrating the relation $l\sim 1/(n_{\rm halo}A_{\rm halo})$ over the redshift distribution of halos along the line of sight.

The equation also shows $S\propto \overline{\sigma^2}_{\rm halo}/(H_0^3 \Omega_b^2 f_{\rm diff}^2 l)$. If we understand more details about the astrophysical properties and distribution of halos, especially the parameters $\overline{\sigma^2}_{\rm halo}$ and $l$, $\sigma_\Delta (S,z)$ can in principle provide a new way to constrain cosmological parameters.

When the redshift is much smaller than 1 ($z\ll 1$), this formula finally returns to the same form as in \citet{2020Nature_Macquart}:
\begin{align}
    {\sigma_\Delta}
    &\sim \frac{\sqrt{S\int_0^z \dd{z'}/E(z')}}{\int_0^z (1+z')\dd{z}/E(z')}\\
    \label{eq:var_delta_F}&\overset{z\ll 1}\sim \frac{\widetilde{F}}{\sqrt{z}}.
\end{align}
However, the standard deviation $\sigma_\Delta$ is not always equal to the parameter $\sigma_{\rm diff}$ \citep{Connor_Sharma, Sharma_2025_ApJ}. Thus we use symbol $\widetilde{F}$ rather than $F$ \citep{2020Nature_Macquart} to distinguish the standard deviation $\sigma_\Delta$ from $\sigma_{\rm diff}$ in $p_{\rm diff}$. Also, because our data reach a maximum redshift of 1.354 for FRB 20230521B, in this work we use the full expression (Eq. \ref{eq:var_delta_S}) numerically integrating for a given redshift and the parameter $S$.

\begin{figure}
\centering
\includegraphics[width=0.5\textwidth]{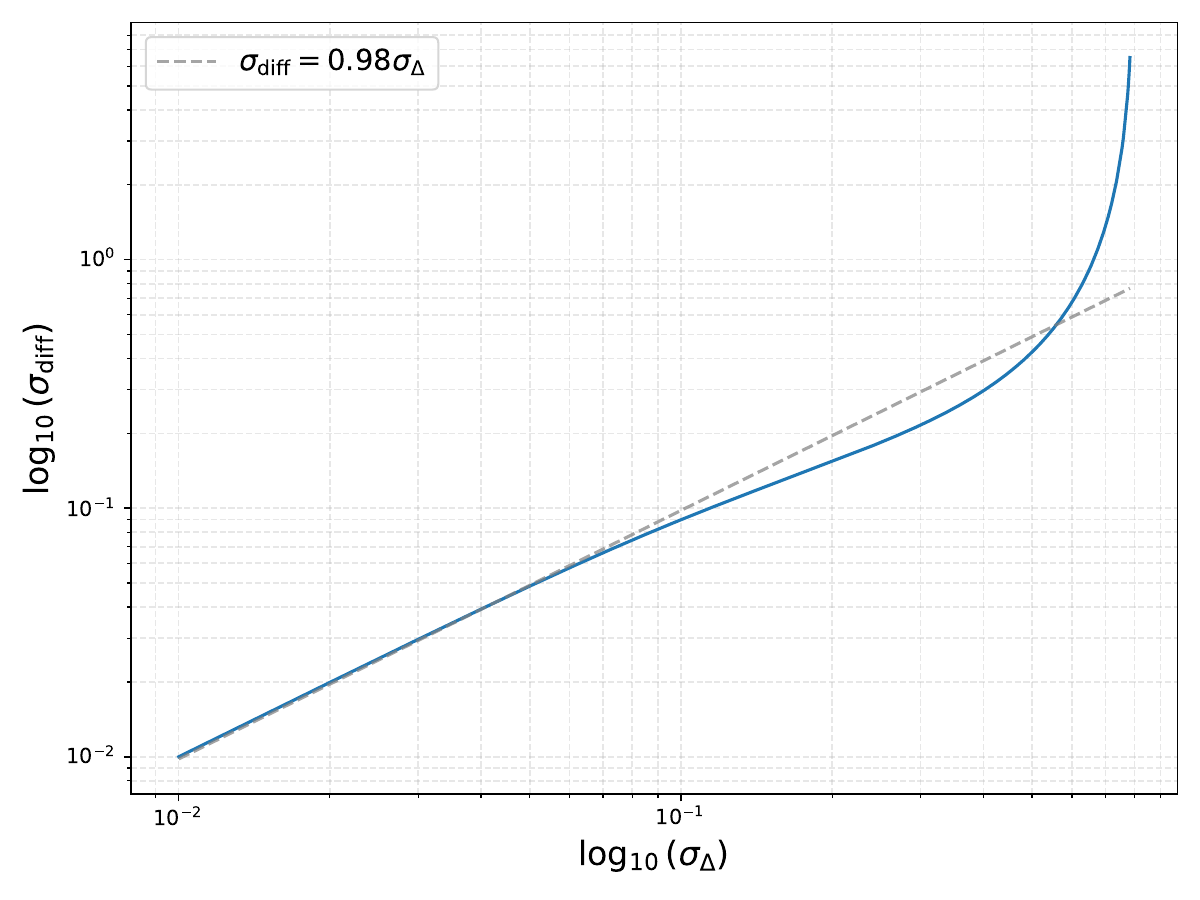}
\caption{$\sigma_\Delta-\sigma_{\rm diff}$ relation in $\log_{10}$ space. The dashed black line is the best linear fitting line for the nearly linear part for $\sigma_{\rm diff}(\sigma_{\Delta})$. }
\label{fig:var_sigma}
\end{figure}

In order to find the relation between the standard deviation $\sigma_\Delta$ and the parameter $\sigma_{\rm diff}$, we derive $\sigma_\Delta$ from the 95\% confidence interval to obtain $\sigma_{\rm diff}$. $\int_{\Delta_1}^{\Delta_2} p_\Delta(\Delta;\sigma_{\rm diff})\dd\Delta =95.45\%$ where $\Delta_2 -\Delta_1=4\sigma_\Delta$. We numerically calculate the $\sigma_\Delta-\sigma_{\rm diff}$ relation and plot it in Fig. \ref{fig:var_sigma} in $\log_{10}$ space. In Fig. \ref{fig:var_sigma}, if $\sigma_\Delta-\sigma_{\rm diff}$ is parallel with $\log_{10}y=\log_{10}x$, because $\sigma_\Delta\sim \widetilde{F}/\sqrt{z}$, $\sigma_{\rm diff}$ can be written as $F/\sqrt{z}$, where $F$ includes some constant times $\widetilde{F}$. However, $\sigma_\Delta$ and $\sigma_{\rm diff}$ only show a linear relationship when $\sigma_\Delta \lesssim 0.60$ or $\sigma_{\rm diff}\lesssim 0.60$. From this region, we obtain the best-fitting function $\log_{10}\sigma_{\rm diff}=\log_{10}\sigma_\Delta \LogKsigmaS$ or $\sigma_{\rm diff}=\KsigmaS\sigma_\Delta$. Recall that when $z\ll 1$ in \citet{2020Nature_Macquart}, $\sigma_{\rm diff}\sim F/\sqrt{z}$. Thus, we have the relation between the two parameters:
\begin{equation}
    F\sim \KsigmaS \widetilde{F}\sim \KsigmaS \sqrt{S}
    \label{eq:F-S}
\end{equation}
In other words, $\sigma_{\rm diff}=F/\sqrt{z}$ is correct only when the redshifts of the FRBs are not too small, within the range
\begin{equation}
    \sigma_{\rm diff}=\frac{F}{\sqrt{z}}\lesssim 0.60 \,\Rightarrow \, z \gtrsim \left(\frac{F}{0.60} \right)^2.
    \label{eq:sigma_linear_range}
\end{equation}
On the other hand, only when $z\ll 1$, we have $\sigma_{\Delta}\sim \widetilde{F}$ or $\sigma_{\rm diff}\sim F/\sqrt{z}$. The two contradictory requirements imply $\widetilde{F}$ or $F$ is not a good approximation. This is also the reason why we use $S$ here. Eq. \ref{eq:F-S} actually only works for a very tight range (depending on the value $S$ or $F$ fit from the data). So Eq. \ref{eq:F-S} is only used to show the equivalent $F$ value derived from $S$, to compare with the $F$ value in previous research works.

For a given search range for $F$, $F/\sqrt{z}$ is only correct when $z$ is not too small; otherwise, very small $z$ will push $\sigma_{\rm diff}$ to grow rapidly and one must numerically consider the $\sigma_\Delta-\sigma_{\rm diff}$ relation. If not, the small redshift will favor a larger $F$ in order to catch up with the real evolution of $\sigma_{\rm diff}$. This might explain why taking the latest FRB observation data would lead to finding a larger value of $F$ on the boundary with a smaller cosmological parameter preference \citep{f_IGM_alpha}.

When $D_c \ll l$ or redshift is very small, the $\sigma_{\rm IGM}$ term dominates (Eq. \ref{eq:var_delta}). This physically corresponds to the case, where the distance to the FRB is very close, so there are no intervening halos in between. Hence, we have:
\begin{align}\label{eq:var_delta_IGM}
    {\sigma_\Delta}
    &\sim\frac{\sigma_{\rm IGM}}{\langle{\DM_{\rm diff} \rangle}}\\
    &\sim \sigma_{\rm IGM} \frac{64\pi G m_p}{21 c H_0 \Omega_b f_d}\frac{1}{\int_0^z (1+z')\dd{z'}/E(z')}\\
    &\overset{z\ll 1}\sim \sigma_{\rm IGM}\frac{\widetilde{F}'}{z} = 0.002\left( \frac{\sigma_{\rm IGM}}{100\ \rm pc\,cm^{-3}}\right)\frac{1}{z}.
\end{align}
Nevertheless, the even faster growth of $\sigma_\Delta(S,z)$ at small redshifts cannot resolve its limit in Fig. \ref{fig:var_sigma}. This logarithmic divergence originates from the definition of the PDF $p_{\rm diff}(\Delta)$. When $z\to 0$, $\langle{\DM_{\rm diff}}\rangle \to 0$ and causes the value of $\Delta$ to fluctuate wildly. This is the reason why $\sigma_\Delta$ increases when $z\to 0$. With a larger standard deviation, the PDF will span a larger domain and $\sigma_{\rm diff}$ will increase accordingly. However, $p_{\rm diff}$ is a quasi-Gaussian distribution multiplied by $\Delta^{-\beta}$. When the parameter $\sigma_{\rm diff}$ in the second quasi-Gaussian distribution term increases, the probability at larger $\Delta$ will be suppressed by $\Delta^{-\beta}$. Thus in Fig. \ref{fig:var_sigma} we see that when $z\to 0$, the parameter $\sigma_{\rm diff}$ needs to grow even faster to make the standard deviation $\sigma_\Delta$ increase. Finally, no matter how fast $\sigma_{\rm diff}$ is growing, $\sigma_\Delta$ does not increase anymore and reaches a limit. On the other hand, when $z$ is large enough, the $\Delta^{-\beta}$ term merely reshapes the PDF and the PDF will be more like a Gaussian distribution. $\sigma_\Delta$ and $\sigma_{\rm diff}$ will be more linear in this case and $\sigma_{\rm diff}$ can be recognized as the ``effective standard deviation''. Overall, the definition of $p_{\rm diff}(\Delta)$ naturally cannot handle the $z\to 0$ case and must have a minimum $z$ limit. The parameters $\alpha$ and $\beta$ will adjust the shape of $p_{\rm diff}(\Delta)$ and affect its applicable redshift range. Given that $\alpha$ and $\beta$ are the parameters describing the gas density profile in the halos, the halo mass function and AGN feedback strength \citep{2025_ZhaoZhang_CROCODILE} will influence the parameters and finally change the applicable redshift range. This fact calls for detailed revisiting of $p_{\rm diff}(\Delta)$, especially $\alpha$ and $\beta$.

A comprehensive consideration of the gas density profile and revisiting the PDF are beyond this project. Hence, to expand the redshift domain where this function applies, in the following analysis we still consider the halo-dominated case in Eq. \ref{eq:var_delta_S}. We neglect the IGM-dominated case in the following sections and hope a more precise PDF at low redshift can use the standard deviation for the IGM term. We also perform a sample cut as mentioned in Section \ref{sec:data}.

\subsubsection{Cosmological dependence of $S$ or $F$}

The derivation of the standard deviation of $\Delta$ reveals the cosmological dependence of the $S$ or $F$ term, usually used in the Macquart relation. For $z \ll 1$, Eq. \ref{eq:var_delta_F} becomes

\begin{equation}
    \sigma_\Delta^2 \sim  \frac{1}{H^{3}_0} \frac{1}{z} \frac{\overline{\sigma^2}_{\rm halo}}{l} \frac{(64\pi G m_p)^2}{c (21 \Omega_b f_{\rm diff})^2},
\end{equation}
with $S$ defined as

\begin{equation}
    S \sim \widetilde{F}^2 \sim \frac{1}{H^{3}_0} \frac{\overline{\sigma^2}_{\rm halo}}{l} \frac{(64\pi G m_p)^2}{c (21 \Omega_b f_{\rm diff})^2},
\end{equation}

For the widely used $F$ parameter, if the $\sqrt{S}-F$ relation is in the linear region, because the factor $0.979$ coincidentally close to $1$, we have:
\begin{equation}
    F^2 \sim S \sim \frac{1}{H^{3}_0} \frac{\overline{\sigma^2}_{\rm halo}}{l} \frac{(64\pi G m_p)^2}{c (21 \Omega_b f_{\rm diff})^2},
\end{equation}
where we confirm the anti-correlation between $F$ and $H_0$ reported in \cite{2024ApJ_Jay_F_H0}. More specifically, we observe that $F \sim H_0^{-3/2}$. By adopting fiducial values for all parameters as in Eq. \ref{eq:var_delta_S}, and $\sqrt{\overline{\sigma^2}_{\rm halo}} \sim 130 {\rm pc/cm^3}$, we provide a helpful relation:

\begin{equation}
    F \simeq 185.75 \left( \frac{H_0}{{\rm km/Mpc \cdot s}} \right)^{-3/2}.
    \label{eq:F_H0}
\end{equation}

\begin{figure}
\centering
\includegraphics[width=0.5\textwidth]{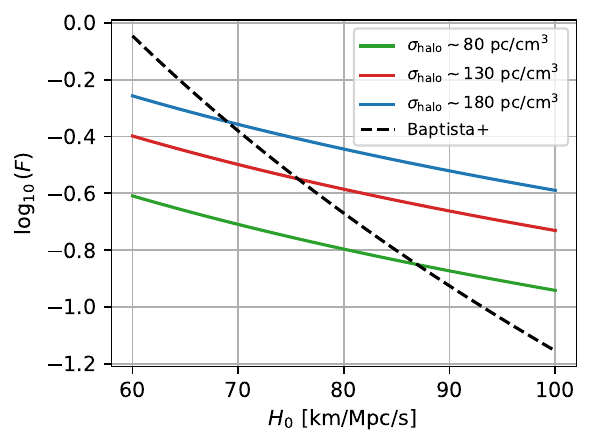}
\caption{Cosmological dependence of $F$ when choosing different value of $\sqrt{\overline{\sigma^2}_{\rm halo}}$.}
\label{fig:F_Ho_dependence}
\end{figure}

This is in good qualitative agreement with \citet[Figure 5]{2024ApJ_Jay_F_H0}, as shown in Fig. \ref{fig:F_Ho_dependence}. The main discrepancy results from the different analyses used: \cite{2024ApJ_Jay_F_H0} show their 2D marginalized distributions, when fitting 78 FRBs with a full model, including $\DM_{\rm host}$. As a result, the relation between $F$ and $H_0$ is affected by correlations between the different parameters. This is in contrast with our relation, which isolates the cosmological dependence of the $F$ term, as a consequence of the halos and IGM modeling of Section \ref{subsubsection:sigma_diff}. In our case, the underlying assumption is that halos dominate the ``effective standard deviation''. The actual $S$ or $F$ will depend on the influence of both halos and the IGM, whose detailed astrophysical modeling is left for a future work. Finally, keeping a more agnostic approach and constraining $S$ and $F$ directly allows a more direct comparison with \cite{2024ApJ_Jay_F_H0}.

\subsection{Likelihood} \label{subsec:likelihood}

Considering the two probability density functions, Eq. \ref{eq:p_host} and Eq. \ref{eq:p_diff}, we reach a modified convolution to calculate the total probability in the DM parameter space for each FRB:
\begin{equation}\label{eq:convolution_pDM_FRB}
  p_{\rm ext}(\DM_{\rm ext}) = \frac{1}
  {\langle{\DM_{\rm diff}}\rangle(z)}
  \int_0^{\DM_{\rm ext}(1+z)} p_{\rm host}(\DM_{\rm host}) p_\Delta \left( \frac{\DM_{\rm FRB}-\DM_{\rm host, z}}
  {\langle{\DM_{\rm diff}}\rangle(z)}
  \right) \dd{\DM_{\rm host}},
\end{equation}
where $\DM_{\rm host,z}=\DM_{\rm host}/(1+z)$. We modify the convolution for two reasons. First, for $p_{\rm host}$ in Eq. \ref{eq:p_host}, the log-normal distribution should be used for all host galaxies in their rest frames. Thus, when we calculate the total $\DM$ or $\DM_{\rm ext}$, $\DM_{\rm host}$ must be divided by the factor $(1+z)$ due to cosmological effects. Second, the variable in $p_{\rm diff}$ in Eq. \ref{eq:p_diff} is actually $\Delta$ rather than $\DM$. In \citet{2020Nature_Macquart}, they partially modified the convolution in their source code, which is slightly different from the equation they used in their paper. We clarify the equations here, to help reproducibility and ease of usage in future research. During the writing of this paper, \citet{2025arXiv_Yuchen_convolution_PDF} also found this modification independently.

For all our well-localized FRB observations, the likelihood in DM parameter space is:
\begin{equation}
    \mathcal{L}(\textbf{DM}_{\rm ext}|\mathbf{H})=\prod_{i=1}^{N_{\rm FRB}}p_{\rm ext, i}(\DM_{\rm ext, i}|\mathbf{H}).
\end{equation}
$\textbf{DM}$ represents the FRB dataset, and $\mathbf{H}$ describes the cosmological parameter space. For example, $\mathbf{H} = \{S, H_0\Omega_b f_{\rm diff}, \sigma_{\rm host}, \mu_{\rm host}\}$. 

For arbitrary $H$ priors, the posterior can then be written as:
\begin{equation}
    p(\mathbf{H}|\textbf{DM}_{\rm ext})=\mathcal{L}(\textbf{DM}_{\rm ext}|\mathbf{H}) p(\mathbf{H})
\end{equation}

In this project, we consider uniform priors for $\mathbf{H}$, and the posterior can be simplified as:
\begin{equation}
    p(\mathbf{H}|\textbf{DM}_{\rm ext}) \propto \mathcal{L}(\textbf{DM}_{\rm ext}|\mathbf{H}).
\end{equation}

The detailed derivation of the likelihood in this section can be found in Appendix \ref{appsec:likelihood}.

\section{Results}\label{sec:results}

In order to show the effect of a more accurate $\sigma_{\rm diff}\left[\sigma_\Delta(S,z)\right]$, we utilize a Markov Chain Monte Carlo (MCMC)  approach to search the parameter space $\mathbf{H}$. We first sample for our $\sigma_{\rm diff}\left[\sigma_\Delta(S,z)\right]$ case. Then, for comparison, we also show the previous $\sigma_{\rm diff}\sim F/\sqrt{z}$ with the same dataset, as well as with all FRBs. We show the prior ranges for each parameter in Table \ref{tab:prior}. For $\exp(\mu)$, we choose a wide range from 10 to 300 as the median value of $\DM_{\rm host}$. For $\sigma_{\rm host}$, we first chose the same range as \citet{2020Nature_Macquart}, but then lowered the maximum limit to save computational time, since it was not affecting the results. $\mathcal{U}[1.0, 5.0]$ is the prior for $H_0 \Omega_b f_{\rm diff}$. This is a wide range based on the fiducial value 2.78 when taking $H_0=67.66\;{\rm km/s/Mpc}$, $\Omega_b=0.04897$ \citep{2020_Planck2018}, and $f_{\rm diff}=0.84$ \citep{2020MNRAS.496L..28L_f_IGM, 2025_ZhaoZhang_CROCODILE}. As for $F$, this is $\mathcal{U}[0.01, 0.5]$ in \citet{2020Nature_Macquart}. We choose the prior for $S$ as $\mathcal{U}[0.008, 0.30]$, based on this range. Finally, we use a prior for $F$ that is wider than before, i.e. $\mathcal{U}[0.01, 1.2]$, because we find a peak at larger values for the latest dataset. Note that both ranges are compatible with the physical estimates we provide in Section \ref{sec:met}.
    
\begin{table}
\centering
\caption{Prior ranges imposed for the parameters in MCMC. }
\label{tab:prior}
\begin{tabular}{l|l}
\hline
\textbf{Parameters} & \textbf{Priors} \\
\hline
$S$ & $\mathcal{U}[0.008, 0.30]$ \\
$F$ & $\mathcal{U}[0.01, 1.2]$ \\
%\hline
$H_0 \Omega_b f_{\rm diff}$ & $\mathcal{U}[1.0, 5.0]$\\
%\hline
$\sigma_{\rm host}$ & $\mathcal{U}[0.2, 1.4]$ \\
%\hline
$ \exp(\mu)$ & $\mathcal{U}[10, 300]$ \\
\hline
\end{tabular}
\end{table}

\subsection{A more accurate $\sigma_{\rm diff}(\sigma_\Delta)$, where $\sigma_\Delta=\sigma_\Delta(S,z)$}\label{subsec:res_our}

\begin{figure}
\centering
\includegraphics[width=1\textwidth]{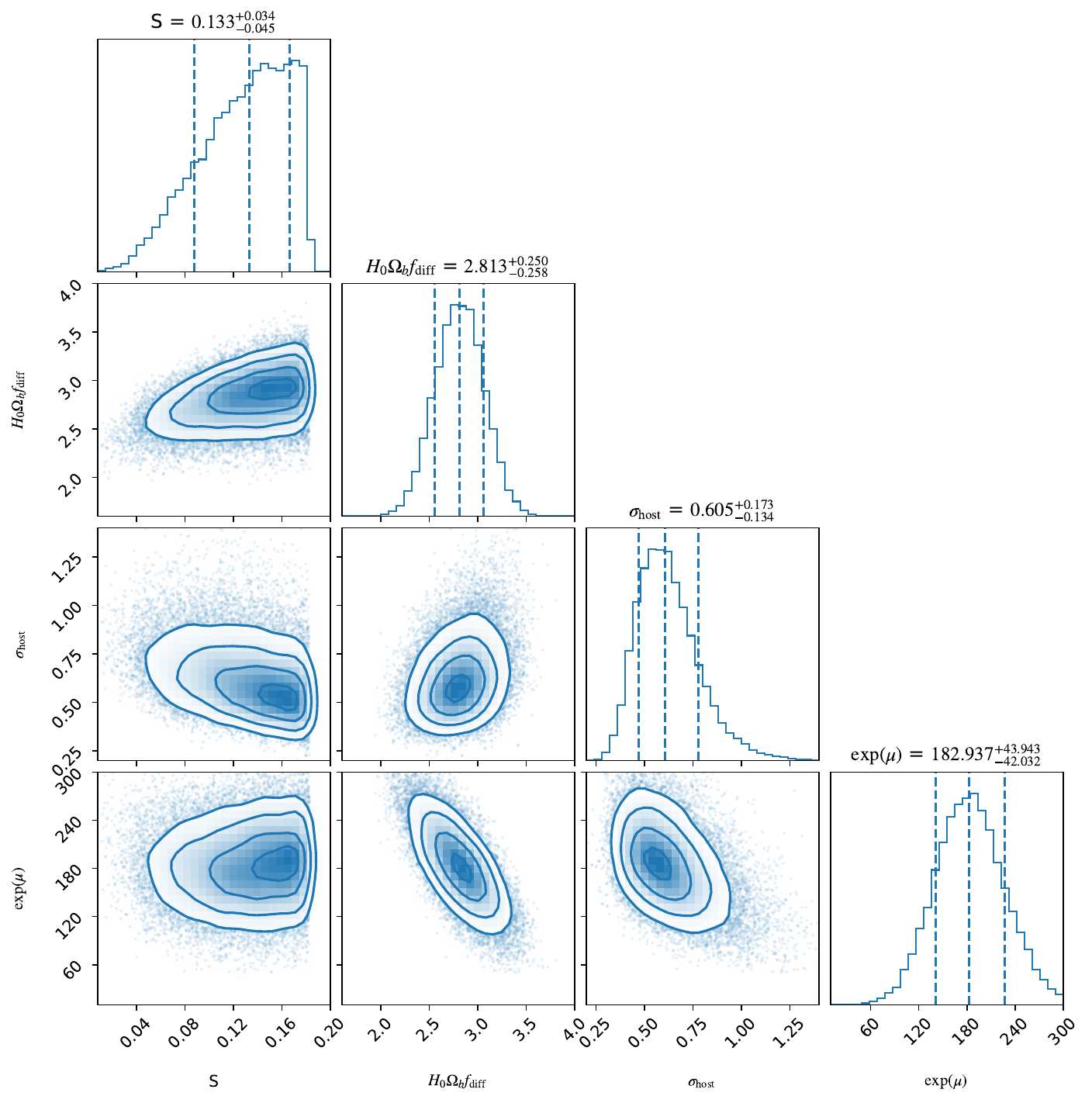}
\caption{MCMC contour plot with our $\sigma_{\rm diff}$ correction}
\label{fig:MCMC}
\end{figure}

First, we show our constraints for parameters $\mathbf{H} = \{S, H_0\Omega_b f_{\rm diff}, \sigma_{\rm host}, \mu_{\rm host}\}$ when considering a more accurate $\sigma_{\rm diff}\left[\sigma_\Delta(S,z)\right]$ and a modified convolution in the likelihood. We assume that $f_{\rm diff}$ is a constant. The three factors in $H_0\Omega_b f_{\rm diff}$ are always bundled and we can only constrain the product. The result of our MCMC inference is shown in Fig. \ref{fig:MCMC}. Note that the prior we choose for $S$ is actually $\mathcal{U}[0.008, 0.30]$. The maximum value for $S$ in Fig. \ref{fig:MCMC} is much smaller than 0.30. The maximum is caused by the rapid growth of $\sigma_{\rm diff}$ for FRBs with small redshift. In our preliminary MCMC runs, we excluded more low-redshift FRBs and confirmed that the redshift cut of $z\geq 0.25$ is sufficient for the posterior distribution to converge to a peak value for $S$ and also utilize the most data.

In Fig. \ref{fig:MCMC}, we obtain $S=\SMCMC$ and $H_0\Omega_b f_{\rm diff}=\HOf\;{\rm km/s/Mpc}$ for the diffuse electron term. Also, we have $\sigma_{\rm host}=\Sigmahost$ and $\exp(\mu)=\Expmu\;{\rm pc/cm^3}$ for the host galaxy term. The lower and upper boundaries show the 68\% confidence level. We list the results in Table \ref{tab:HOf}. If we combine the CMB \citep{2020_Planck2018} measurement $\Omega_b h^2=0.0224$ and take $f_{\rm diff}=\fdiff$ (based on some earlier results including \citet{2012ApJ...759...23S_f_IGM} ($f_{\rm diff}=0.82$), \citet{2020MNRAS.496L..28L_f_IGM} ($f_{\rm diff}=0.84$), \citet{2022ApJ...931...88C_f_IGM} ($f_{\rm diff}=0.85$), \citet{1998ApJ...503..518F_f_IGM} ($f_{\rm diff}=0.83$), \citet{2025_ZhaoZhang_CROCODILE} ($f_{\rm diff}=0.84$)), we can obtain the $H_0=\Hubble\;{\rm km/s/Mpc}$. For the parameter $S$, although it differs from the previously widely used parameter $F$, we use Eq. \ref{eq:F-S} to estimate the equivalent $F$ when considering that $\sigma_{\Delta}(S,z)\sim \widetilde{F}/\sqrt{z}$ is still linear with $\sigma_{\rm diff}$. The result shows the equivalent $F=\F$. This is consistent with the 5 FRBs ``golden sample'' result in \citet{2020Nature_Macquart}.
% This is consistent with ``5 golden samples'' result in \citet{2020Nature_Macquart}.

Recall that for the linear range for $\sigma_{\rm diff}-\sigma_\Delta$, where $F$ parameterization is valid in Eq. \ref{eq:sigma_linear_range}, we have:
\begin{equation}
    \sigma_{\rm diff}\sim \frac{F}{\sqrt{z}}\lesssim 0.6
\end{equation}
Combining with the $F=\F$ constraint, the $F$ parameter actually works only for
\begin{equation}
    z\gtrsim 0.354
\end{equation}
Moreover, $\widetilde{F}$ or $F$ are approximated from $S$ when $z\ll1$ (see Eq. \ref{eq:var_delta_F} or \citet{2020Nature_Macquart}). Thus the $F$ parameterization is plausible only under conflicting assumptions. For a rigorous estimation, the $F$ parameter should be used for
\begin{equation}
    0.354 \lesssim z < 1.
\end{equation}

\subsection{A comparison with $\sigma_{\rm diff}\sim Fz^{-1/2}$}

\begin{figure}[h]
\centering
\includegraphics[width=1\textwidth]{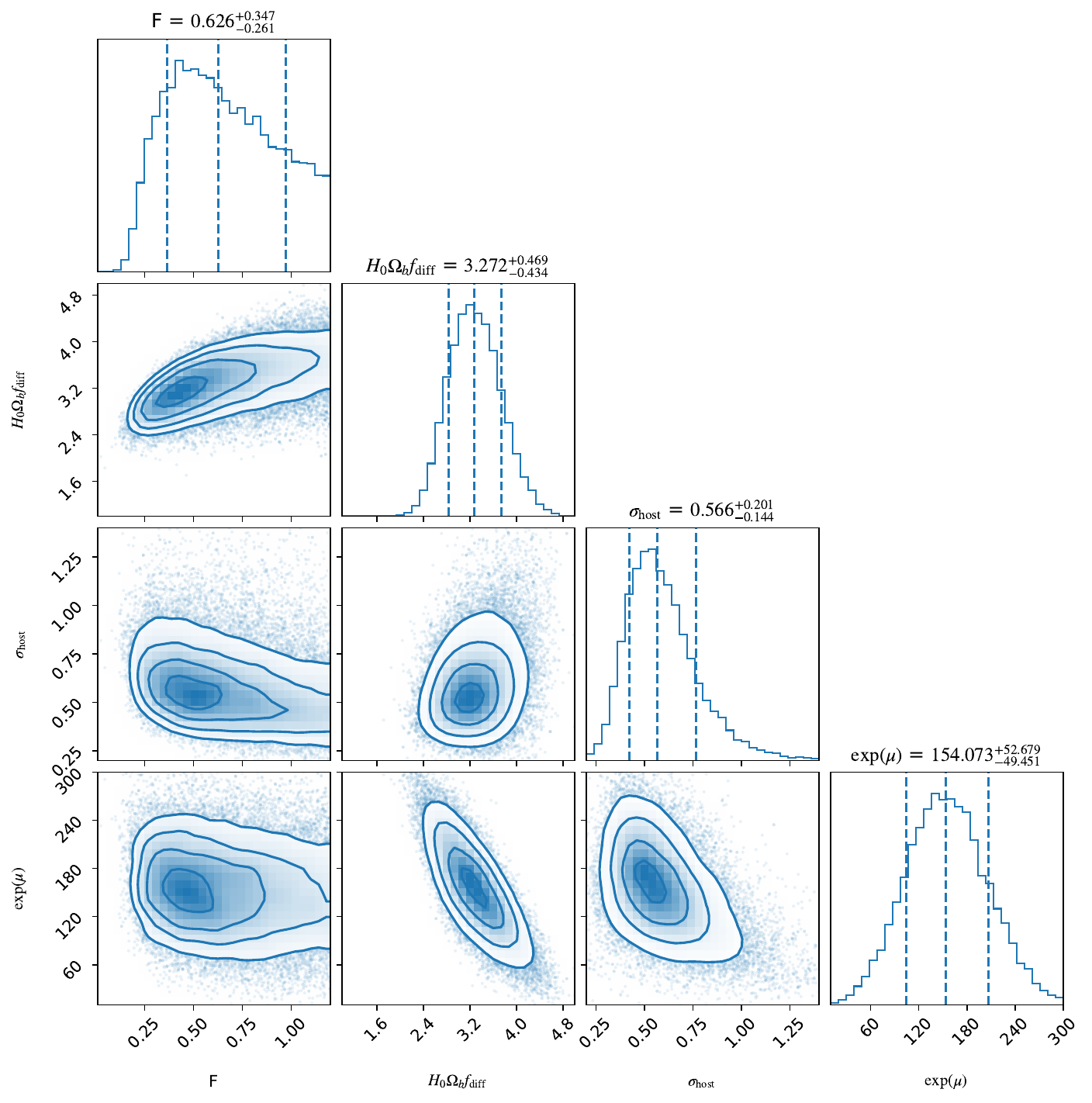}
\caption{Macquart's $\sigma_{\rm diff}=F/\sqrt{z}$ with FRBs $z\ge 0.25$}
\label{fig:Macquart03}
\end{figure}

To ensure that the differences are due to the more accurate treatment to calculate $\sigma_{\rm diff}$ rather than a dataset bias, we also run an MCMC with the widely used $\sigma_{\rm diff}\sim Fz^{-1/2}$ approximation, taking the same dataset while dropping the small-redshift samples. The result in Fig. \ref{fig:Macquart03} is more similar to the 5 FRBs ``golden sample'' result in \citet{2020Nature_Macquart}. However, the $H_0 \Omega_b f_{\rm diff}$ is much larger, corresponding to a very small Hubble constant $H_0=57.506_{-7.209}^{+8.794}\;{\rm km/s/Mpc}$ if taking $\Omega_b h^2=0.0224$ \citep{2020_Planck2018} and $f_{\rm diff}=0.84$ \citep{2020MNRAS.496L..28L_f_IGM, 2025_ZhaoZhang_CROCODILE}. In this method, $\sigma_{\rm diff}\sim F/\sqrt{z}\sim 1.252$ is much larger than the $\sigma_{\rm diff}-\sigma_\Delta$ linear region, which cause errors in $F$ and $H_0 \Omega_b f_{\rm diff}$.

\begin{figure}
\centering
\includegraphics[width=1\textwidth]{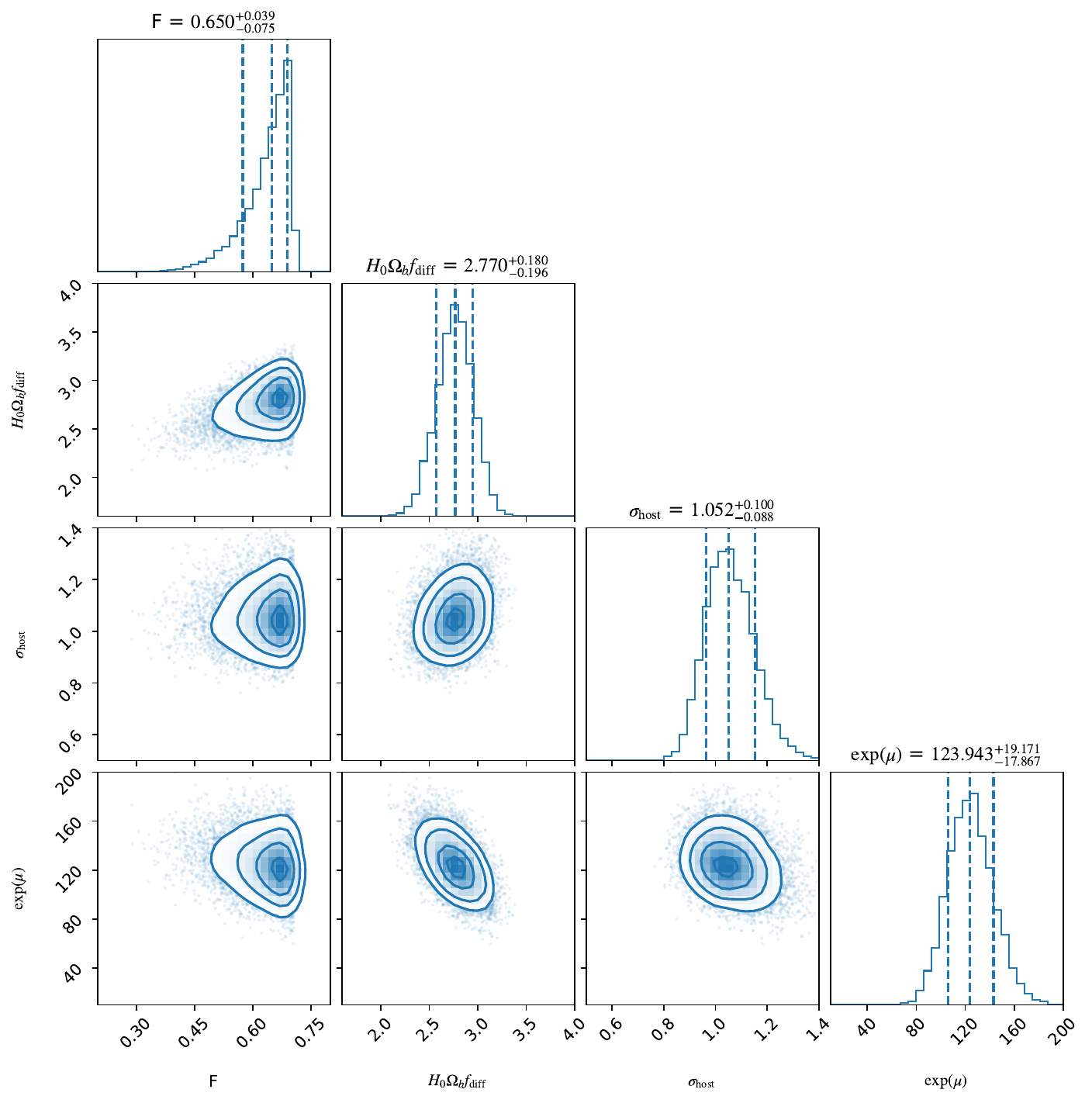}
\caption{Macquart's $\sigma_{\rm diff}=F/\sqrt{z}$ with 115 localized FRBs}
\label{fig:Macquart}
\end{figure}

In Fig. \ref{fig:Macquart}, we also show the result when we take \citet{2020Nature_Macquart}'s original $\sigma_{\rm diff}\sim Fz^{-1/2}$ approximation and the entire FRB sample without dropping any low-redshift FRBs except FRB 20190520B and FRB 20220831A. In this case, $F$ exhibits rapid growth and gets cut off. When we slightly drop some FRBs with very small redshifts, we get closer to the peak of $F$ but also gain a larger $H_0\Omega_bf_{\rm diff}$, and finally get the result in Fig. \ref{fig:Macquart03} when dropping FRBs with $z<0.25$. This also implies that $\sigma_{\rm diff}\sim Fz^{-1/2}$ and $p_{\rm diff}(\Delta)$ fail to describe the PDF in the nearby universe. The limit in $\sigma_\Delta(S,z)$ actually provides a way to estimate the redshift range ($0.354 \lesssim z < 1$) that $p_{\rm diff}$ is accurate and where $\sigma_{\rm diff}\sim Fz^{-1/2}$ is useful (when $\sigma_\Delta-\sigma_{\rm diff}$ is linear).

\begin{table}
\centering
\caption{Results for $\mathbf{H} = \{S$ (or $F$ if \citet{2020Nature_Macquart} original approximation)$, H_0\Omega_b f_{\rm diff}, \sigma_{\rm host}, \mu_{\rm host}\}$ with 68\% confidence interval. We show the results of our correction using MCMC. We also show \citet{2020Nature_Macquart}'s original method using the same FRB samples and 115 localized FRBs for comparison. We also show the corresponding $F$ (or $S$) according to Eq. \ref{eq:F-S} to easily show their magnitudes. The $H_0$ column is calculated from $H_0 \Omega_b f_{\rm diff}$ using $\Omega_b h^2=0.0224$ \citep{2020_Planck2018} and $f_{\rm diff}=0.84$ \citep{2020MNRAS.496L..28L_f_IGM, 2025_ZhaoZhang_CROCODILE}.}
\label{tab:HOf}
\begin{tabular}{!{\vrule width 1pt}l!{\vrule width 1pt}l!{\vrule width 1pt}l|l!{\vrule width 1pt}l!{\vrule width 1pt}l!{\vrule width 1pt}l!{\vrule width 1pt}l!{\vrule width 1pt}}
\hline
\textbf{Method} & \textbf{Data} & $S$ & $F$ & $H_0 \Omega_b f_{\rm diff}$ & $\sigma_{\rm host}$ & $\exp(\mu) $ & $H_0$ \\
 & & & & \scriptsize(km/s/Mpc) & & \scriptsize(pc/${\rm cm^3}$) & \scriptsize(km/s/Mpc) \\
\hline
$\sigma_{\rm diff}(S,z)$ & $z\geq0.25$ & $\SMCMC$ & $\sim \F$ & $\HOf$ & $\Sigmahost$ & $\Expmu$ & $\Hubble$ \\
$\sigma_{\rm diff}=F/\sqrt{z}$ & $z\geq0.25$ & $\sim \MacScut$ & $0.626_{-0.261}^{+0.347}$ & $3.272_{-0.434}^{+0.469}$ & $0.566_{-0.144}^{+0.201} $ & $154.072_{-49.451}^{+52.679}$ & $57.506^{+8.794}_{-7.209}$\\
$\sigma_{\rm diff}=F/\sqrt{z}$ & 115 FRBs samples  & $> \MacS$ &  $> 0.650_{-0.075}^{+0.039}$ & $>2.770_{-0.196}^{+0.180}$ & $1.052_{-0.088}^{+0.100}$ & $123.943_{-17.867}^{+19.171}$ & $<67.928_{-4.145}^{+5.172}$ \\
\hline
\end{tabular}
\end{table}

\section{Summary and discussion}\label{sec:conclusion}

In this paper, we write down a clearer form for the FRB $\DM$ likelihood used to obtain cosmological constraints and discuss the effect of an accurate approach to derive $\sigma_{\rm diff}$ in $p_{\rm diff}$. More specifically, by exploiting the actual PDF of $\DM_{\rm diff}$, we find the relation of its true standard deviation with the effective $\sigma_{\rm diff}$ used in the literature $\sigma_{\rm diff}=\sigma_{\rm diff}\left[\sigma_\Delta(S,z)\right]$. We use MCMC to perform parameter estimation for $\mathbf{H} = \{S, H_0\Omega_b f_{\rm diff}, \sigma_{\rm host}, \mu_{\rm host}\}$ {and investigate the performance of cosmological constraints with an improved physical modeling. Additionally, we find} a suitable redshift range for $p_{\rm diff}$. In summary, we reach the following conclusions:

\begin{itemize}

\item $\sigma_{\rm diff}$ in $p_{\rm diff}(\Delta)$ is not the true standard deviation of the variable $\Delta$. Only when the standard deviation is not too large or the redshift is not too small, can $\sigma_\Delta$ and $\sigma_{\rm diff}$ take a linear relation and $\sigma_{\rm diff}$ serve as an ``effective standard deviation''. From our MCMC result combining the equivalent $F\sim \F$, with the linear limit $\sigma_{\rm diff}\sim F/\sqrt{z}\lesssim 0.60$, we conclude that only for $z\gtrsim 0.354$, $\sigma_{\rm diff}-\sigma_\Delta$ has a linear relation. However, only when $z\ll 1$ one can approximate $\sqrt{S}\sim F$ and have $\sigma_{\rm diff}\sim \sigma_\Delta\propto \sqrt{D_c}/
{\langle{\DM_{\rm diff}}\rangle}
\propto \widetilde{F}/\sqrt{z}$. This reveals that the widely used $F$ parameter in $\sigma_{\rm diff}\sim F/\sqrt{z}$ only works under two contrived and contradictory assumptions. Otherwise, we must take the accurate form $\sigma_{\rm diff}\left[\sigma_\Delta (S,z)\right]$. 

\item In the non-linear regime for $\log(\sigma_\Delta)-\log(\sigma_{\rm diff})$, $\sigma_\Delta$ has a maximum limit. When $\sigma_\Delta$ approaches the maximum value, $\sigma_{\rm diff}$ diverges logarithmically. Very large $\sigma_\Delta$ values are associated with very small redshifts, which means that $p_{\rm diff} (\Delta)$ may not work in the low-redshift regime, which reminds us that $\sigma_{\rm diff}$ cannot be an ``effective standard deviation'' there. In the future, $p_{\rm diff}$ still needs to be revisited in more detail, especially in the low-redshift regime. One way is to consider the halo and IGM components at different redshifts. The $\alpha$ and $\beta$ in $p_{\rm diff}(\Delta)$ may need to be modified when we also consider the IGM component in the diffuse electron term. AGN feedback will also affect the gas density profile in the halos and influence $\alpha$ and $\beta$. Another possible approach is to directly take the PDF for $\DM_{\rm diff}$ rather than for $\Delta$. This intrinsic PDF will have a convergent standard deviation. However, $p_{\rm diff}$ will require a wider search range to find $C_0$, which will cause other numerical problems. 

\item  With our more accurate parameter consideration, $S$, $\widetilde{F}$, or $F$ actually acquire a physical meaning (see Eq. \ref{eq:var_delta_S}). For the first time we provide a physical explanation for the correlation observed in \citep{2024ApJ_Jay_F_H0}. This allows us to propose a new approach to constrain cosmological parameters. These parameters represent the combination of some cosmological parameters, halo baryon fluctuations, and the mean observed halo path from the FRB perspective. If we understand more about the halo density profiles and baryon distributions, these parameters provide another way to measure $1/S\propto H_0^3 \Omega_b^2 f_{\rm diff}^2$ and to constrain cosmological parameters.

\item  With our new $\sigma_{\rm diff}\left[\sigma_\Delta(S,z)\right]$, we obtain the constraints $S=\SMCMC$, $H_0\Omega_b f_{\rm diff}=\HOf\;{\rm km/s/Mpc}$, and $\widetilde{\DM}_{\rm host}=\exp(\mu)=\Expmu\;{\rm pc/cm^3}$. When combined with the Planck18 results \citep{2020_Planck2018} and $f_{\rm diff}=\fdiff$, we constrain the Hubble constant to $H_0=\Hubble\;{\rm km/s/Mpc}$. Many previous cosmological constraints from other methods are within this range. One can expect FRBs to become an accurate probe to constrain cosmological parameters if, in the future, we have more localized FRBs and understand more about their host galaxies as well as the intervening halos.

\end{itemize}

\begin{acknowledgments}

We thank J. X. Prochaska for the informative discussion to allow us know about the details of the treatment in \cite{2020Nature_Macquart}. We also acknowledge some helpful suggestions from Carl-Johan Haster. JZ thanks the helpful discussion with Yuanhong Qu, Zi-Liang Zhang, Yihan Wang and Rui-Chong Hu. JZ and BZ's work is supported by the Nevada Center for Astrophysics, NASA 80NSSC23M0104 and a Top Tier Doctoral Graduate Research Assistantship (TTDGRA) at University of Nevada, Las Vegas.

\end{acknowledgments}

\vspace{5mm}
% \facilities{UNLV}

\software{NumPy \citep{Numpy_harris2020array}, astropy \citep{2013A&A...558A..33A,2018AJ....156..123A}, Matplotlib \citep{Hunter:2007_matplotlib}, SciPy \citep{2020SciPy-NMeth}, pandas \citep{mckinney-proc-scipy-2010_pandas, 2023zndo...7857418T_pandas2.0.1}, emcee \citep{emcee}, PyGEDM \citep{PyGEDM}. 
The core pipeline and reproduction codes for the cosmological analysis in this work are publicly available on GitHub (\url{https://github.com/MariosNT/FRB_GW_association_cosmo}) and permanently archived on Zenodo via \dataset[doi:10.5281/zenodo.20282512]{https://doi.org/10.5281/zenodo.20282512} \citep{Code_zenodo}.
}

\appendix

\section{Likelihood calculation for DM of FRBs}
\label{appsec:likelihood}

\subsection{2 PDFs case}

In this section, we provide a detailed derivation of Eq. \ref{eq:convolution_pDM_FRB} that calculates the likelihood of the observed DM ($\DM_{\rm obs}$) of a given FRB with redshift $z$. In statistics, assume the sum of two independent, random variables $X, Y$ with probability density functions (PDF) $p_X(x)$, $p_Y(y)$. The random variable $Z$ calculated by $Z=X+Y$, would have a PDF given from the convolution of $p_X$ and $p_Y$ \citep[e.g.][]{Intro_to_Prob_2019}:
\begin{equation}\label{eq:ap_convolution}
    p_{Z}(z)= (p_X * p_Y)(z) = \int_{a}^{b}p_X(\tau)p_Y(z-\tau)d{\tau}.
\end{equation}
The upper and lower limits may be adjusted according to the range of the variables.

For an FRB, we subtract the MW term from its $\DM_{\rm obs}$ and get extragalactic DM ($\DM_{\rm ext}$). The remaining terms include the diffuse medium term and the host environment term:
\begin{equation}
    \DM_{\rm ext} = \DM_{\rm diff}+\DM_{\rm host,z}.
\end{equation}
Note that the $\DM$ that we actually measure is in the Earth frame, meaning that a redshift correction is applied to the $\DM_{\rm host,z}$, i.e. $\DM_{\rm host, z} = \DM_{\rm host}/(1+z)$, where $z=z_{\rm FRB}$. This means that in reality, we want to calculate the PDF of:
\begin{equation}
    \DM_{\rm ext}  = \DM_{\rm diff}+\DM_{\rm host, z}= \DM_{\rm diff}+\frac{\DM_{\rm host}}{1+z}.
\end{equation}
At the same time, one must be careful because in Eq. \ref{eq:p_diff}, $p_{\rm diff}(\Delta=\DM/{\langle{\DM_{\rm diff}}\rangle}
)$ is not in the same variable space as $p_{\rm host}(\DM)$. This means that a change of variables is required for both distributions. Demanding the conservation of probability, we have:
\begin{align}
    \label{eq:ap_var_transformation_1}
    & p_{\DM_{\rm host,z}}(\DM_{\rm host, z})\dd{\DM_{\rm host, z}} = p_{\rm host}(\DM_{\rm host})\dd{\DM_{\rm host} }
    & \Rightarrow & \quad p_{\DM_{\rm host,z}}(\DM_{\rm host, z}) = p_{\rm host}(\DM_{\rm host})(1+z)\\
    \label{eq:ap_var_transformation_2}
    & p_{\DM_{\rm diff}}(\DM_{\rm diff})\dd{\DM_{\rm diff}} = p_{\Delta}(\Delta)\dd{\Delta }
    & \Rightarrow & \quad p_{\DM_{\rm diff}}(\DM_{\rm diff}) = p_\Delta(\Delta)/\langle \DM_{\rm diff}\rangle. 
\end{align}

Then, for convolution in the $\DM$ space, for an individual FRB $\DM_{\rm FRB} = \DM_{\rm host, z} + \DM_{\rm IGM}$, we have
\begin{align}\label{eq:ap_pDM_FRB_convo}
    p(\DM_{\rm FRB}) &= (p_{\DM_{\rm host, z}} * p_{\DM_{\rm diff}}) = \int_0^{\DM_{\rm FRB}} p_{\DM_{\rm host, z}}(\DM_{\rm host, z}) p_{\DM_{\rm IGM}} (\DM_{\rm FRB}-\DM_{\rm host, z})\dd\DM_{\rm host, z} \\ \nonumber
    &= \int_0^{\DM_{\rm FRB}(1+z)} p_{\rm host}(\DM_{\rm host}) p_{\Delta} \left( \frac{\DM_{\rm FRB}-\DM_{\rm host, z}}{
    \langle{\DM_{\rm diff}}\rangle(z)
    } \right) \frac{\dd\DM_{\rm host}}{
    \langle{\DM_{\rm diff}}\rangle
    }, \nonumber
\end{align}
where we have used Eq. (\ref{eq:ap_var_transformation_1}, \ref{eq:ap_var_transformation_2}) in the second step. More clearly, one has
\begin{equation}\label{eq:ap_pDM_FRB}
  \boxed{p(\DM_{\rm FRB}) = 
  \frac{1}{
  {\langle{\DM_{\rm diff}}\rangle}
  }
  \int_0^{\DM_{\rm FRB}(1+z)} 
  p_{\rm host}(\DM_{\rm host}) 
  p_{\Delta} \left( \frac{\DM_{\rm FRB}-\DM_{\rm host,z}}{{\langle{\DM_{\rm diff}}\rangle}
  } \right) \dd\DM_{\rm host}}
\end{equation}

Alternatively, if one does convolution in the $\Delta$ space, one has
\begin{equation}
    \DM_{\rm FRB} = \DM_{\rm host, z} + \DM_{\rm diff} = \DM_{\rm host, z} +  \Delta \cdot {\langle{\DM_{\rm diff}}\rangle}
    \Rightarrow \DM_{\rm host}/(1+z) = \DM_{\rm FRB}-\Delta \cdot 
    {\langle{\DM_{\rm diff}}\rangle}.
\end{equation}
The PDF becomes
\begin{align}
    p(\DM_{\rm FRB}) &= (p_{\DM_{\rm host, z}} * p_{\DM_{\rm diff}}) = \int_0^{\DM_{\rm FRB}} p_{\DM_{\rm host, z}}(\DM_{\rm FRB} - \DM_{\rm diff}) p_{\DM_{\rm diff}} (\DM_{\rm diff})d\DM_{\rm diff} \nonumber \\
    &= \int_0^{\DM_{\rm FRB}/{\langle{\DM_{\rm diff}}\rangle(z)}
    } p_{\rm host, z}(\DM_{\rm FRB} - \Delta \cdot {\langle{\DM_{\rm diff}}\rangle}
    ) p_{\Delta} (\Delta)d\Delta  \nonumber \\
    &= \int_0^{\DM_{\rm FRB}/{\langle{\DM_{\rm diff}}\rangle(z)}
    } (1+z)p_{\rm host} \left((1+z) \cdot (\DM_{\rm FRB} - \Delta \cdot {\langle{\DM_{\rm diff}}\rangle}
    )\right) p_{\Delta} (\Delta)d\Delta,
\end{align}
which finally gives
\begin{equation}\label{eq:ap_pDelta_FRB}
  \boxed{p(\DM_{\rm FRB}) = (1+z) \int_0^{\DM_{\rm FRB}/{\langle{\DM_{\rm diff}}\rangle(z)}
  } p_{\rm host} \left[ (1+z) \cdot (\DM_{\rm FRB} - \Delta \cdot {\langle{\DM_{\rm diff}}\rangle}
  ) \right] p_{\Delta} (\Delta)d\Delta}
\end{equation}
Obviously Eq. \ref{eq:ap_pDelta_FRB} and Eq. \ref{eq:ap_pDM_FRB} are equivalent.

For multiple, well-localized FRB observations, one has
\begin{equation}
    \mathcal{L}(\textbf{DM}|\mathbf{H})=\prod_{i=1}^{N_{\rm FRB}}p_{\DM_{\rm FRB, i}}(\DM_{\rm FRB, i}|\mathbf{H}).
\end{equation}
\textbf{DM} represents the FRBs dataset, and $\mathbf{H}$ describes the parameter space. 

The final posterior can then be written as:
\begin{equation}
    p(\mathbf{H}|\textbf{DM}) \propto \mathcal{L}(\textbf{DM}|\mathbf{H}) \cdot p(\mathbf{H}).
\end{equation}

\subsection{Likelihood based on conditional probabilities}

In the previous subsection, we derived the convolution for FRBs PDFs used in in this project, but we have skipped some steps. The combined PDF is actually a conditional probability. In this subsection, we derive it mathematically. 

To calculate the probability of $Z=X+Y$, we start by calculating the joint probability $p(Z, X, Y)$:
\begin{equation}
    p(Z, X, Y) = p(Z|X, Y)p(X,Y) = p(Z|X, Y)p(X|Y)p(Y) = p(Z|X, Y)p(X)p(Y),
\end{equation}
where in the first step, we used the definition of the conditional probability, and in the second, we used the assumption that $X, Y$ are independent.

To get $p(Z)$ we marginalise the probability over $X, Y$:
\begin{equation}
    p(Z) = \int dX \int dY \ p(Z, X, Y) = \int dX \int dY \ p(Z|X, Y)p(X)p(Y).
\end{equation}
Since $Z=X+Y$, the variables are not independent and if we know $X, Y$, then there is only one possible value of $Z$, i.e.
\begin{equation}
    p(Z|X, Y) = \delta \left(Z-(X+Y)\right),
\end{equation}
where $\delta$ is the Dirac $\delta$-function.

Inserting this in the integral, we get:
\begin{equation}
    p(Z) = \int dX \int dY \ \delta \left(Z-(X+Y)\right)p(X)p(Y) = \int dX \ p(X) p(Z-X),
\end{equation}
so we end up with the convolution as in Eq. \ref{eq:ap_convolution}. For $Z=\DM_{\rm FRB}, X=\DM_{\rm host, z}$ and $Y=\DM_{\rm diff}$, this yields:
\begin{equation}
   p(\DM_{\rm FRB}) = \int_0^{\DM_{\rm FRB}} p_{\DM_{\rm host, z}}(\DM_{\rm host, z}) p_{\DM_{\rm diff}} (\DM_{\rm FRB}-\DM_{\rm host, z})d\DM_{\rm host, z}
\end{equation}
which is the same as Eq. \ref{eq:ap_pDM_FRB_convo}, and one can then follow a similar procedure as above.

\subsection{The case of three PDFs}

In reality, FRB $\DM$ is not just composed of the MW, diffuse medium, and host galaxy terms. For example, \citet{2024_Surajit} also considered a Gaussian distribution for the Milky Way halo component. We consider one extra term: the third component in the external DM term, i.e. $\DM_{\rm src}$, which refers to the DM contribution from the immediate environment of the FRB source. Then one defines the total host component as $\DM_{\rm hosts}$:
\begin{equation}
    \DM_{\rm hosts}=\DM_{\rm host}+\DM_{\rm src}
\end{equation}
It is easy to get the probability for the host term, i.e.
\begin{equation}
    p_{\rm hosts}(\DM_{\rm hosts})=
    \int_{0}^{\DM_{\rm hosts}} 
    p_{\rm host}(\DM_{\rm hosts}-\DM_{\rm src})p_{\rm src}(\DM_{\rm src})
    \dd{\DM_{\rm src}}
\end{equation}

One can simply replace $\DM_{\rm host}$ with $\DM_{\rm hosts}$ in Eq. \ref{eq:ap_pDM_FRB} to obtain
\begin{equation}
\begin{aligned}
  & p(\DM_{\rm FRB}) \\ & = 
  \frac{1}{{\langle{\DM_{\rm diff}}\rangle(z)}
  }
  \int_{0}^{\DM_{\rm FRB}(1+z)}  
  p_{\Delta} \left( \frac{\DM_{\rm FRB}-\DM_{\rm hosts,z}}{\langle{\DM_{\rm diff}}\rangle(z)
  } \right) 
  \int_{0}^{\DM_{\rm hosts}} p_{\rm host}(\DM_{\rm hosts}-\DM_{\rm src})p_{\rm src}(\DM_{\rm src})\dd{\DM_{\rm src}}
  \dd{\DM_{\rm hosts}}
\end{aligned}
\end{equation}
It is straightforward to show that this is equivalent to
\begin{equation}\label{eq:ap_pDM_FRB3}
\begin{aligned}
  & p(\DM_{\rm FRB}) \\ &=
  \frac{1}{{\langle{\DM_{\rm diff}}\rangle(z)}
  }\int_{0}^{\DM_{\rm host}=\DM_{\rm FRB}(1+z)}  
  \int_{0}^{\DM_{\rm src}=\DM_{\rm FRB}(1+z)-\DM_{\rm host}}
  p_{\Delta} \left( \frac{\DM_{\rm FRB}-\DM_{\rm host,z}-\DM_{\rm src,z}}
  {\langle{\DM_{\rm diff}}\rangle(z)}
  \right)  p_{\rm host}(\DM_{\rm host})\\
  & \times
  p_{\rm src}(\DM_{\rm src})
  \dd{\DM_{\rm src}}
  \dd{\DM_{\rm host}}.
\end{aligned}
\end{equation}
This is just an example for the three PDFs. One can modify the treatment to include other $\DM $ contributions as required.

\section{FRB Data}\label{appsec:FRBs}

\startlongtable
\begin{deluxetable}{cccccccc}
\label{tab:data}
\centering
\tablecaption{Well localized FRB samples}
\tablehead{\colhead{FRB} & \colhead{z} & \colhead{gl} & \colhead{gb} & \colhead{$\DM_{\rm obs}$} & \colhead{$\DM_{\rm MW, ISM}$(NE2001)} & \colhead{$\DM_{\rm MW, ISM}$(YMW16)} & \colhead{Ref.} \\
& & (deg) & (deg) & $ (\rm pc \, cm^{-3})$ & $(\rm pc \, cm^{-3})$ & $(\rm pc \, cm^{-3})$ & 
}
\startdata
FRB 20121102A & 0.19273 & 174.89 & -0.23 & 557.0 & 188.42 & 287.07 & 1, 40 \\
FRB 20171020A & 0.00867 & 29.3 & -51.3 & 114.1 & 38.37 & 25.84 & 36, 13 \\
FRB 20180301A & 0.3304 & 204.412 & -6.481 & 522.0 & 150.73 & 252.56 & 24 \\
FRB 20180814A & 0.068 & 136.46 & 16.58 & 190.9 & 87.83 & 108.41 & 37 \\
FRB 20180916B & 0.0338 & 129.71 & 3.73 & 348.76 & 198.91 & 324.82 & 5, 41, 34 \\
FRB 20180924B & 0.3214 & 0.7424 & -49.4147 & 361.75 & 40.5 & 27.65 & 3, 6, 23, 11, 20, 42 \\
FRB 20181030A & 0.00385 & 133.4 & 40.9 & 103.5 & 40.35 & 32.24 & 27 \\
FRB 20181112A & 0.4755 & 342.5995 & -47.6988 & 589.0 & 41.72 & 29.03 & 4, 6, 23, 20, 42 \\
FRB 20181220A & 0.02746 & 105.24 & -10.73 & 209.4 & 125.85 & 122.13 & 28 \\
FRB 20181223C & 0.03024 & 207.75 & 79.51 & 112.5 & 19.93 & 19.15 & 28 \\
FRB 20190102C & 0.29 & 312.6537 & -33.4931 & 364.55 & 57.4 & 43.28 & 6, 23, 20, 42 \\
FRB 20190110C & 0.12244 & 65.6 & 42.1 & 221.6 & 37.07 & 29.95 & 33 \\
FRB 20190303A & 0.064 & 97.5 & 65.7 & 222.4 & 29.67 & 21.83 & 37 \\
FRB 20190418A & 0.07132 & 179.3 & -22.93 & 184.5 & 70.11 & 85.6 & 28 \\
FRB 20190425A & 0.03122 & 42.06 & 33.02 & 128.2 & 48.8 & 38.81 & 28 \\
FRB 20190520B & 0.241 & 359.67 & 29.91 & 1204.7 & 60.21 & 50.23 & 9, 11 \\
FRB 20190523A & 0.66 & 117.03 & 44.0 & 760.8 & 37.18 & 29.88 & 2 \\
FRB 20190608B & 0.11778 & 53.2088 & -48.5296 & 339.5 & 37.27 & 26.62 & 6, 23, 11, 20, 42 \\
FRB 20190611B & 0.3778 & 312.9352 & -33.2818 & 322.4 & 57.83 & 43.67 & 32, 6, 20, 42 \\
FRB 20190614D & 0.6 & 136.3 & 16.5 & 959.2 & 88.0 & 109.06 & 35 \\
FRB 20190711A & 0.522 & 310.9081 & -33.902 & 592.0 & 56.49 & 42.61 & 32, 6, 20, 42 \\
FRB 20190714A & 0.2365 & 289.6972 & 48.9359 & 504.7 & 38.49 & 31.16 & 32, 20, 42 \\
FRB 20191001A & 0.23 & 341.2267 & -44.9039 & 507.0 & 44.17 & 31.08 & 32, 20, 42 \\
FRB 20191106C & 0.10775 & 105.7 & 73.2 & 332.2 & 25.01 & 20.54 & 33 \\
FRB 20191228A & 0.2432 & 20.5553 & -64.9245 & 297.0 & 32.95 & 19.92 & 24, 20, 42 \\
FRB 20200120E & 0.00014 & 142.19 & 41.22 & 87.82 & 40.67 & 32.23 & 26, 31 \\
FRB 20200223B & 0.06024 & 118.1 & -33.9 & 201.8 & 45.53 & 36.99 & 33 \\
FRB 20200430A & 0.1608 & 17.1396 & 52.503 & 380.1 & 27.18 & 26.08 & 32, 20, 42 \\
FRB 20200906A & 0.3688 & 202.257 & -49.9989 & 577.8 & 35.84 & 37.87 & 24, 20, 42 \\
FRB 20201123A & 0.0507 & 340.23 & -9.68 & 433.55 & 251.66 & 162.66 & 7 \\
FRB 20201124A & 0.098 & 177.6 & -8.5 & 413.52 & 140.13 & 196.69 & 8 \\
FRB 20210117A & 0.214 & 45.9175 & -57.6464 & 729.2 & 34.38 & 23.1 & 25, 11, 20, 42 \\
FRB 20210320C & 0.28 & 318.8729 & 45.3081 & 384.6 & 39.29 & 30.39 & 11, 20, 42 \\
FRB 20210405I & 0.066 & 338.19 & -4.59 & 565.17 & 516.78 & 349.21 & 16 \\
FRB 20210410D & 0.1415 & 312.32 & -34.13 & 578.78 & 56.19 & 42.24 & 12, 11 \\
FRB 20210603A & 0.1772 & 119.71 & -41.58 & 500.147 & 39.53 & 30.79 & 18 \\
FRB 20210807D & 0.12927 & 39.8612 & -14.8775 & 251.9 & 121.17 & 93.66 & 11, 20 \\
FRB 20211127I & 0.046946 & 312.0214 & 43.5427 & 234.83 & 42.47 & 31.46 & 11, 31, 20, 42 \\
FRB 20211203C & 0.3439 & 314.5185 & 30.4361 & 636.2 & 63.73 & 48.38 & 11, 20, 42 \\
FRB 20211212A & 0.0707 & 244.0081 & 47.3154 & 200.0 & 38.75 & 27.46 & 11, 20, 42 \\
FRB 20220105A & 0.2785 & 18.555 & 74.808 & 583.0 & 22.04 & 20.63 & 11, 20, 42 \\
FRB 20220204A & 0.4012 & 102.26 & 27.06 & 612.2 & 52.84 & 48.7 & 15, 38, 30 \\
FRB 20220207C & 0.0433 & 106.94 & 18.39 & 262.3 & 76.1 & 83.27 & 39, 15, 38, 30 \\
FRB 20220208A & 0.351 & 107.62 & 15.36 & 437.0 & 90.4 & 107.92 & 15, 38, 30 \\
FRB 20220307B & 0.2481 & 116.24 & 10.47 & 499.15 & 128.25 & 186.98 & 39, 15, 38, 30 \\
FRB 20220310F & 0.478 & 140.02 & 34.8 & 462.15 & 46.34 & 39.51 & 39, 15, 38, 30 \\
FRB 20220319D & 0.0112 & 129.18 & 9.11 & 110.95 & 139.71 & 210.96 & 39, 15, 38, 30, 19 \\
FRB 20220330D & 0.3714 & 134.18 & 42.93 & 468.1 & 38.98 & 30.63 & 15, 38, 30 \\
FRB 20220418A & 0.6213 & 110.75 & 44.47 & 623.45 & 36.65 & 29.54 & 39, 15, 38, 30 \\
FRB 20220501C & 0.381 & 11.1777 & -71.4731 & 449.5 & 30.62 & 14.0 & 20, 38, 42 \\
FRB 20220506D & 0.3004 & 108.35 & 16.51 & 369.93 & 84.58 & 97.69 & 39, 15, 38, 30 \\
FRB 20220509G & 0.0894 & 100.94 & 25.48 & 269.5 & 55.6 & 52.06 & 39, 15, 38, 30 \\
FRB 20220529A & 0.1839 & 130.7877 & -41.858 & 246.0 & 39.95 & 30.92 & 22 \\
FRB 20220610A & 1.015 & 8.8392 & -70.1857 & 1458.1 & 30.96 & 13.58 & 20, 42 \\
FRB 20220717A & 0.36295 & 19.8352 & -17.632 & 637.34 & 118.33 & 83.22 & 17 \\
FRB 20220725A & 0.1962 & 0.0017 & -71.1863 & 290.1 & 30.73 & 11.58 & 20, 42 \\
FRB 20220726A & 0.3619 & 139.97 & 17.57 & 686.55 & 83.65 & 101.14 & 15, 38, 30 \\
FRB 20220825A & 0.2414 & 106.99 & 17.79 & 651.2 & 78.46 & 86.91 & 39, 15, 38, 30 \\
FRB 20220831A & 0.262 & 110.96 & 12.47 & 1146.25 & 110.21 & 147.77 & 30 \\
FRB 20220912A & 0.0771 & 347.27 & 48.7 & 219.46 & 32.5 & 28.6 & 10 \\
FRB 20220914A & 0.1139 & 104.31 & 26.13 & 631.05 & 54.68 & 51.11 & 15, 38, 30 \\
FRB 20220918A & 0.491 & 300.6851 & -46.2342 & 643.0 & 40.74 & 28.87 & 20, 42 \\
FRB 20220920A & 0.1582 & 104.92 & 38.89 & 315.0 & 39.86 & 33.36 & 39, 15, 38, 30 \\
FRB 20221012A & 0.2847 & 101.14 & 26.14 & 442.2 & 54.35 & 50.55 & 39, 15, 38, 30 \\
FRB 20221027A & 0.5422 & 142.66 & 33.96 & 452.5 & 47.62 & 41.07 & 15, 38, 30 \\
FRB 20221029A & 0.975 & 140.39 & 38.01 & 1391.05 & 43.15 & 35.42 & 15, 38, 30 \\
FRB 20221101B & 0.2395 & 113.25 & 11.06 & 490.7 & 122.68 & 174.2 & 15, 38, 30 \\
FRB 20221106A & 0.2044 & 220.901 & -50.8788 & 343.2 & 34.79 & 31.84 & 20, 42 \\
FRB 20221113A & 0.2505 & 139.53 & 16.99 & 411.4 & 86.31 & 105.52 & 38, 30 \\
FRB 20221116A & 0.2764 & 124.47 & 8.71 & 640.6 & 145.05 & 223.48 & 38, 30, 20 \\
FRB 20221219A & 0.553 & 103.19 & 34.07 & 706.708 & 43.97 & 37.98 & 38, 30 \\
FRB 20230124A & 0.0939 & 107.55 & 40.25 & 590.574 & 39.13 & 32.36 & 38, 30 \\
FRB 20230203A & 0.1464 & 188.7125 & 54.087 & 420.1 & 36.26 & 22.95 & 21 \\
FRB 20230216A & 0.531 & 242.6 & 46.33 & 828.0 & 39.46 & 28.15 & 38, 30 \\
FRB 20230222A & 0.1223 & 204.7164 & 8.6957 & 706.1 & 134.2 & 188.08 & 21 \\
FRB 20230222B & 0.11 & 49.6176 & 49.9787 & 187.8 & 27.73 & 26.29 & 21 \\
FRB 20230307A & 0.2706 & 127.35 & 45.0 & 608.854 & 37.4 & 29.26 & 38, 30 \\
FRB 20230311A & 0.1918 & 157.7134 & 16.0395 & 364.3 & 92.46 & 115.68 & 21 \\
FRB 20230501A & 0.3015 & 112.43 & 11.51 & 532.471 & 118.55 & 165.22 & 38, 30 \\
FRB 20230521B & 1.354 & 115.65 & 9.97 & 1342.9 & 133.4 & 197.8 & 30 \\
FRB 20230526A & 0.157 & 290.171 & -63.4721 & 316.2 & 31.86 & 21.86 & 20, 42 \\
FRB 20230626A & 0.327 & 105.68 & 38.34 & 451.2 & 40.21 & 33.85 & 38, 30 \\
FRB 20230628A & 0.127 & 135.48 & 42.21 & 344.952 & 39.59 & 31.17 & 38, 30 \\
FRB 20230703A & 0.1184 & 137.2098 & 67.475 & 291.3 & 26.93 & 20.67 & 21 \\
FRB 20230708A & 0.105 & 342.6288 & -33.3877 & 411.51 & 60.33 & 43.99 & 20, 42 \\
FRB 20230712A & 0.4525 & 132.31 & 43.69 & 587.567 & 38.44 & 30.1 & 38, 30 \\
FRB 20230718A & 0.035 & 259.4629 & -0.3666 & 476.67 & 420.65 & 449.99 & 20, 42 \\
FRB 20230730A & 0.2115 & 158.8166 & -17.8164 & 312.5 & 85.16 & 97.38 & 21 \\
FRB 20230814A & 0.553 & 112.56 & 13.2 & 696.4 & 104.76 & 137.79 & 30 \\
FRB 20230814B & 0.5535 & 111.25 & 12.26 & 696.4 & 111.87 & 151.38 & 30 \\
FRB 20230902A & 0.3619 & 256.9906 & -53.3387 & 440.1 & 34.14 & 25.53 & 20, 42 \\
FRB 20230926A & 0.0553 & 68.2353 & 27.484 & 222.8 & 52.62 & 43.72 & 21 \\
FRB 20231005A & 0.0713 & 57.1653 & 44.3532 & 189.4 & 33.47 & 28.79 & 21 \\
FRB 20231011A & 0.0783 & 127.2287 & -20.943 & 186.3 & 70.35 & 65.69 & 21 \\
FRB 20231017A & 0.245 & 100.6098 & -21.6519 & 344.2 & 64.55 & 55.64 & 21 \\
FRB 20231025B & 0.3238 & 93.4332 & 29.434 & 368.7 & 48.59 & 43.37 & 21 \\
FRB 20231120A & 0.0368 & 140.43 & 37.94 & 438.9 & 43.22 & 35.5 & 38, 30 \\
FRB 20231123A & 0.0729 & 199.3094 & -15.7427 & 302.1 & 89.75 & 136.86 & 21 \\
FRB 20231123B & 0.2621 & 105.71 & 38.38 & 396.857 & 40.36 & 33.82 & 38, 30 \\
FRB 20231128A & 0.1079 & 105.6852 & 73.224 & 331.6 & 25.01 & 20.54 & 21 \\
FRB 20231201A & 0.1119 & 163.0405 & -22.7702 & 169.4 & 69.95 & 74.72 & 21 \\
FRB 20231204A & 0.0644 & 97.6238 & 65.9281 & 221.0 & 29.83 & 21.79 & 21 \\
FRB 20231206A & 0.0659 & 161.0582 & 27.4819 & 457.7 & 59.14 & 59.29 & 21 \\
FRB 20231220A & 0.3355 & 143.29 & 31.69 & 491.2 & 50.27 & 45.17 & 30 \\
FRB 20231223C & 0.1059 & 52.3107 & 32.0802 & 165.8 & 47.87 & 38.64 & 21 \\
FRB 20231226A & 0.1569 & 236.5768 & 48.6458 & 329.9 & 38.09 & 26.69 & 20, 42 \\
FRB 20231229A & 0.019 & 135.3449 & -26.4433 & 198.5 & 58.18 & 51.77 & 21 \\
FRB 20231230A & 0.0298 & 195.8701 & -25.2387 & 131.4 & 61.58 & 83.27 & 21 \\
FRB 20240114A & 0.1306 & 57.73 & -31.68 & 527.7 & 49.42 & 38.77 & 14, 29 \\
FRB 20240119A & 0.376 & 112.37 & 43.24 & 483.1 & 37.36 & 30.32 & 30 \\
FRB 20240123A & 0.968 & 138.14 & 15.34 & 1462.0 & 94.02 & 119.71 & 30 \\
FRB 20240201A & 0.042729 & 222.1335 & 47.9692 & 374.5 & 38.62 & 29.14 & 20, 42 \\
FRB 20240210A & 0.023686 & 14.4396 & -86.2116 & 283.73 & 28.69 & 17.9 & 20, 42 \\
FRB 20240213A & 0.1185 & 136.49 & 41.57 & 357.4 & 40.1 & 31.76 & 30 \\
FRB 20240215A & 0.21 & 102.29 & 30.17 & 549.5 & 48.29 & 43.11 & 30 \\
FRB 20240229A & 0.287 & 131.13 & 44.11 & 491.15 & 37.94 & 29.81 & 30 \\
FRB 20240310A & 0.127 & 291.7066 & -72.2713 & 601.8 & 30.1 & 19.83 & 20, 42 \\
\enddata
\tablecomments{Reference: (1) \citet{2017Nature_121102}; (2) \citet{2019Nature_Ravi}; (3) \citet{2019Science180924}; (4) \citet{2019Science181112}; (5) \citet{2020Nature_180916}; (6) \citet{2020Nature_Macquart}; (7) \citet{2022MNRAS_Rajwade}; (8) \citet{2022MNRAS_Ravi}; (9) \citet{2022Nature_CHNiu}; (10) \citet{2023ApJ_20220912A}; (11) \citet{2023ApJ_Garden}; (12) \citet{2023MNRAS.524.2064C}; (13) \citet{2023PASA_20171020A_revisited}; (14) \citet{2024ApJ_20240114A}; (15) \citet{2024ApJ_25}; (16) \citet{2024MNRAS20210405I}; (17) \citet{2024MNRAS_2trb_sys}; (18) \citet{2024NatAs...8.1429C}; (19) \citet{2025AJ_20220319D}; (20) \citet{2025ASKAP}; (21) \citet{2025arXiv_CHIME_localized}; (22) \citet{2025arXiv_YeLi}; (23) \citet{Bhandari_2020}; (24) \citet{Bhandari_2022}; (25) \citet{Bhandari_2023}; (26) \citet{Bhardwaj_2021}; (27) \citet{Bhardwaj_2021_181030A}; (28) \citet{Bhardwaj_2024}; (29) \citet{Chen_2025}; (30) \citet{Connor_Sharma}; (31) \citet{Glowacki_2023}; (32) \citet{Heintz_2020}; (33) \citet{Ibik_2024}; (34) \citet{Kaur_2022_180916}; (35) \citet{Law_2020}; (36) \citet{Mahony_2018}; (37) \citet{Michilli_2023}; (38) \citet{Sharma}; (39) \citet{Sherman_2023}; (40) \citet{Tendulkar_2017_121102}; (41) \citet{Tendulkar_2021_180916}; (42) \citet{highres_ASKAP_arxiv2025}}
\end{deluxetable}

\bibliography{ref}{}
\bibliographystyle{aasjournal}
\end{CJK*}
\end{document}